\documentclass{aa}
\input psfig.tex
\usepackage{graphicx}
\usepackage{natbib}
\bibpunct{(}{)}{;}{a}{}{,}

\usepackage{array}
\usepackage{graphics}
\usepackage{latexsym}
\usepackage{amssymb}
\usepackage{amsmath}
\usepackage{fancyhdr}
\usepackage{float}
\usepackage{morefloats}
\usepackage{slashbox}
\usepackage{multirow}
\usepackage[toc,page]{appendix}
\bibpunct{(}{)}{;}{a}{}{,}
\include{hyphe}

\usepackage[english]{babel}

\begin{document}
\title{T-PHOT\thanks{\textsc{t-phot} is publicly available for downloading from \textit{www.astrodeep.eu/t-phot/} .}: A new code for PSF-matched, prior-based, 
multiwavelength extragalactic deconfusion photometry}

\author{E. ~Merlin \inst{1}
\and A. ~Fontana \inst{1}
\and H. ~C. ~Ferguson \inst{2}
\and J. ~S. ~Dunlop \inst{3}
\and D. ~Elbaz \inst{4}
\and N. ~Bourne \inst{3}
\and V. ~A. ~Bruce \inst{3}
\and F. ~Buitrago \inst{3,10,11}
\and M. ~Castellano \inst{1}
\and C. ~Schreiber \inst{4}
\and A. ~Grazian \inst{1}
\and R. ~J. ~McLure \inst{3}
\and K. ~Okumura \inst{4}
\and X. ~Shu  \inst{4,8} 
\and T. ~Wang  \inst{4,9}
\and R. ~Amor\'{i}n \inst{1}
\and K. ~Boutsia \inst{1}
\and N. ~Cappelluti \inst{5}
\and A. ~Comastri \inst{5}
\and S. ~Derriere \inst{6}
\and S. ~M. ~Faber \inst{7}
\and P. ~Santini \inst{1} 
}

\institute{INAF - Osservatorio Astronomico di Roma, Via Frascati 33, I - 00040 Monte Porzio Catone (RM), Italy \email{emiliano.merlin\char64oa-roma.inaf.it}\label{inst1}
\and Space Telescope Science Institute, 3700 San Martin Drive, Baltimore, MD 21218, USA \label{inst2}
\and SUPA\thanks{Scottish Universities Physics Alliance}, Institute for Astronomy, University of Edinburgh, Royal Observatory, Edinburgh, EH9 3HJ, U.K. \label{inst3}
\and Laboratoire AIM-Paris-Saclay, CEA/DSM/Irfu - CNRS - Universit\'e Paris Diderot, CEA-Saclay, pt courrier 131, F-91191 Gif-sur-Yvette, France \label{inst4}
\and INAF - Osservatorio Astronomico di Bologna, Via Ranzani 1, I - 40127, Bologna, Italy \label{inst5}
\and Observatoire astronomique de Strasbourg, Université de Strasbourg, CNRS, UMR 7550, 11 rue de l’Université, F-67000 Strasbourg, France \label{inst6}
\and UCO/Lick Observatory, University of California, 1156 High Street, Santa Cruz, CA 95064, USA \label{inst7}
\and Department of Physics, Anhui Normal University, Wuhu, Anhui, 241000, China\label{inst8}
\and School of Astronomy and Astrophysics, Nanjing University, Nanjing, 210093, China\label{ints9}
\and Instituto de Astrof\'{\i}sica e Ci\^{e}ncias do Espa\c{c}o, Universidade de Lisboa, OAL, Tapada da Ajuda, PT1349-018 Lisbon, Portugal\label{inst10}
\and Departamento de F\`{i}sica, Faculdade de Ci\^{e}ncias, Universidade de Lisboa, Edif\`{i}cio C8, Campo Grande, PT1749-016 Lisbon, Portugal \label{inst11}\\
}

\authorrunning{E. Merlin et al.}
\titlerunning{\textsc{t-phot}}

\abstract{The advent of deep multiwavelength extragalactic surveys 
has led to the necessity for advanced and fast methods for photometric 
analysis. In fact, codes which allow analyses of the same regions of the sky 
observed at different wavelengths and resolutions are becoming essential 
to thoroughly exploit current and future data. In this context,
a key issue is the \emph{confusion} (i.e. blending) of sources
in low-resolution images.}
{We present \textsc{t-phot}, a publicly available 
software package developed within the \textsc{astrodeep} 
project. \textsc{t-phot} is aimed at extracting  
accurate photometry from low-resolution images, where the blending 
of sources can be a serious problem for the accurate and unbiased
measurement of fluxes and colours.}
{\textsc{t-phot} can be considered as the next generation to 
\textsc{tfit}, providing significant improvements over and above 
it and other similar codes (e.g. \textsc{convphot}).
\textsc{t-phot} gathers data from a high-resolution 
image of a region of the sky, and uses this information (source 
positions and morphologies) to obtain priors for the photometric 
analysis of the lower resolution image of the same field. \textsc{t-phot}
can handle different types of datasets as input priors, namely
i) a list of objects that will be used to obtain cutouts from the 
real high-resolution image; ii) a set of analytical models (as 
\texttt{.fits} stamps); iii) a list of unresolved, point-like sources, 
useful for example for far infrared wavelength domains.}
{By means of simulations and analysis of real datasets,
we show that \textsc{t-phot} yields accurate estimations 
of fluxes within the intrinsic uncertainties of the method, 
when systematic errors are taken into account (which can be 
done thanks to a flagging code given in the output).
\textsc{t-phot} is many times faster than similar
codes like \textsc{tfit} and \textsc{convphot} (up to hundreds, 
depending on the problem and the method adopted), whilst
at the same time being more robust and more versatile. This makes 
it an excellent choice for the analysis of large datasets. 
When used with the same parameter sets as for \textsc{tfit}
it yields almost identical results (although in a much shorter time); 
in addition we show how the use of different settings 
and methods significantly enhances the performance.}
{\textsc{t-phot} proves to be a state-of-the-art 
tool for multiwavelength optical to far-infrared image photometry. 
Given its versatility and robustness, \textsc{t-phot} 
can be considered the preferred choice for combined photometric 
analysis of current and forthcoming extragalactic 
imaging surveys.}

\keywords{Galaxy, photometry, multiwavelength, software}
\maketitle

\section{Introduction}

\begin{figure*}[ht] 
\includegraphics[width=7cm]{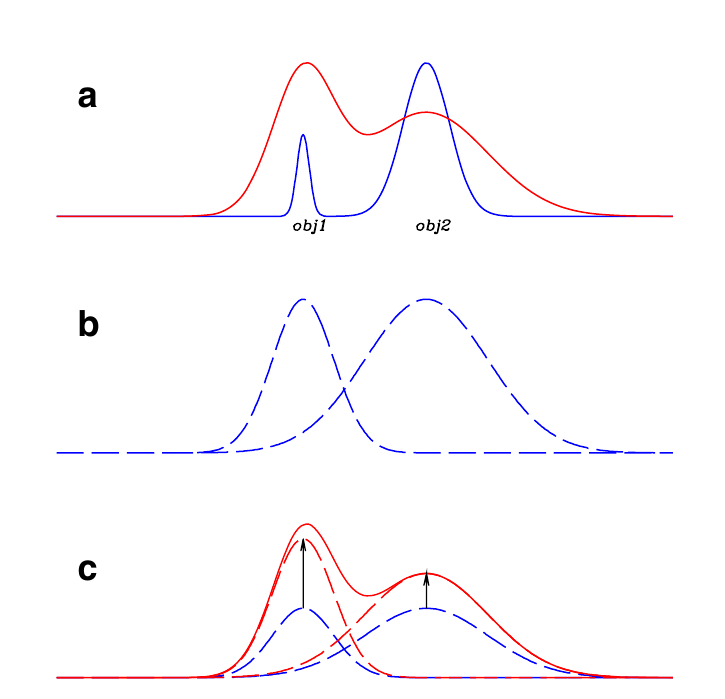}
\centering
\caption{Schematic representation of the PSF-matched algorithm
implemented in \textsc{t-phot}. Top: two objects are clearly detected
and separated in the high-resolution detection image (blue line). The same two 
objects are blended in the low-resolution measurement image and have different
colours (red line). Middle: the two objects are isolated in the detection image
and are individually smoothed to the PSF of the measurement image, to obtain
normalized model templates. Bottom: the intensity of each object is scaled 
to match the global profile of the measurement image. The scaling factors
are found with a global $\chi^2$ minimization over the object areas.
Image from \citet{DeSantis2007}.}\label{method}
\end{figure*}

Combining observational data from the same regions of the sky 
in different wavelength domains has become common practice 
in the past few years
\citep[e.g.][ and many others]{Agueros2005, Obric2006, Grogin2011}.
However, the use of both space-based and ground-based imaging 
instruments, with different sensitivities, pixel scales, 
angular resolutions, and survey depths, raises a number 
of challenging difficulties in the data analysis process.

In this context, it is of particular interest to obtain detailed 
photometric measurements for high-redshift galaxies in the 
near-infrared (NIR; corresponding to rest-frame optical)
and far-infrared (FIR) domains.
In particular, great attention must be paid to bandpasses 
containing spectral features which allow a thorough
investigation of the sources, disentangling degenerate 
observational features, and obtaining crucial
clues to the understanding of the galactic physics
\citep[e.g.][]{Daddi2004, Fontana2009}. 
At $z>3$, for example, photometry longward of $H$-band is needed
to locate and measure the size of the Balmer break. 
A passive galaxy at $z \simeq 6$ (with
the Balmer break lying longward of the $K$-band) can have 
$H$-band and $3.6 \mu$m fluxes compatible, for example, with a 
star forming, dusty galaxy at $z \simeq 1$, and $K-$band photometry
is necessary in order to disentangle the degeneracy. However,
the limited resolution of the ground based $K-$band observations
can impose severe limits on the reliability of traditional
aperture or even Point Spread Function (PSF) fitting photometry. 
In addition, IRAC photometry is of crucial importance 
so that reliable photometric redshifts of red and high-$z$ 
sources can be obtained, and robust stellar mass estimates can be
derived.

To address this, a high-resolution image (HRI), for example obtained 
from the Hubble Space Telescope in the optical domain, 
can be used to retrieve detailed information on the positions 
and morphologies of the sources in a given region of the sky.
Such information can be subsequently used to perform the photometric
analysis of the lower resolution image (LRI), using the HRI data as priors.
However, simply performing aperture photometry on the LRI at the positions 
measured in the HRI can be dramatically affected by neighbour 
contamination for reasonably sized apertures. On the other hand, performing 
source extraction on both images and matching the resulting catalogues is 
compromised by the inability to deblend neighbouring objects, and may
introduce significant inaccuracies in the cross-correlation process. 
PSF-matching techniques that degrade high-resolution data to match the 
low-resolution data discard much of the valuable information obtained 
in the HRI, reducing all images to the ``lowest common 
denominator'' of angular resolution. Moreover, crowded-field, PSF-fitting 
photometry packages such as \textsc{daophot} \citep{Stetson1987} 
perform well if the sources in the LRI are unresolved, but are unsuitable 
for analysis of even marginally resolved images of extragalactic sources.

A more viable approach consists of taking advantage of the 
morphological information given by the HRI, in order to obtain high-resolution
cutouts or models of the sources. These priors can then be degraded to
the resolution of the LRI using a suitable \textit{convolution kernel},
constructed by matching the PSFs of the HRI and of the LRI. Such 
low-resolution templates, normalized to unit flux, can then be placed 
at the positions given by the HRI detections, and the multiplicative 
factor that must be assigned to each model to match the measured flux 
in each pixel of the LRI will give the measured flux of that source.
Such an approach, although relying on some demanding assumptions as
described in the following sections, has proven to be 
efficient. It has been implemented in such public codes as \textsc{tfit}
\citep{Laidler2007} and \textsc{convphot} \citep{DeSantis2007},
and has already been utilized successfully in previous 
studies \citep[e.g.][]{Guo2013, Galametz2013}.

In this paper we describe a new software package, \textsc{t-phot}, 
developed at INAF-OAR as part of the \textsc{astrodeep} 
project\footnote{\textsc{astrodeep} is a coordinated and comprehensive 
program of i) algorithm/software development and testing; 
ii) data reduction/release, and iii) scientific data 
validation/analysis of the deepest multi-wavelength cosmic surveys.
For more information, visit \textit{http://astrodeep.eu} .}. 
The \textsc{t-phot} software can be considered a new, largely 
improved version of \textsc{tfit}, supplemented with many 
of the features of \textsc{convphot}.
Moreover, it adds many important new options, including
the possibility of adopting different types of priors (namely
real images, analytical models, or point-sources). In
particular, it is possible to use \textsc{t-phot} on
FIR and sub-millimetric (sub-mm) datasets, as a competitive alternative to 
the existing dedicated software such as \textsc{FastPhot} 
\citep{Bethermin2010} and \textsc{DesPhot} \citep{Roseboom2010,Wang2014}.
This makes  \textsc{t-phot} a versatile tool, suitable
for the photometric analysis of a very broad range of
wavelengths from UV to sub-mm.

\textsc{t-phot} is a robust and easy-to-handle code, with a 
precise structural architecture (a \textsc{Python} envelope 
calling \texttt{C/C++} core codes) in which different routines 
are encapsulated, implementing various numerical/conceptual methods, 
to be chosen by simple switches in a parameter file. While a standard
default ``best choice'' mode is provided and suggested, 
the user is allowed to select a preferred setting.

One of the main advantages of \textsc{t-phot} is a significant saving 
of computational time with respect to both \textsc{tfit} and 
\textsc{convphot} (see Sect. \ref{times}). This has been
achieved with the use of fast \texttt{C} modules 
and an efficient structural arrangement of the code. 
In addition to this, we demonstrate how different choices 
of parameters influence the performace, and can be optimized 
to significantly improve the final results with respect to 
\textsc{tfit}, for example.

The plan of the paper is as follows. Sect. \ref{code} provides
a general introduction to the code, its mode of operation and
its algorithms. In Section \ref{limits} we discuss some assumptions, 
limitations and caveats of the method.
Section \ref{validation} presents a comprehensive
set of tests, based on simulated and real datasets,
to assess the performance of the code and to fully illustrate its
capabilities and limitations. Section \ref{times} briefly
discusses the computational performances of \textsc{t-phot}
and provides some reference computational timescales. 
Finally, in Section \ref{conclusions} the key features of \textsc{t-phot} 
are summarized, and outstanding issues and potential complications
are briefly discussed.

\section{General description of the code} \label{code}

As described above, \textsc{t-phot} uses spatial and morphological
information gathered from a HRI to measure the fluxes in a LRI. 
To this end, a linear system is built and solved via matricial
computing, minimizing the $\chi^2$ (in which the numerically 
determined fluxes for each detected source are compared to the 
measured fluxes in the LRI, summing the contributions of all pixels). 
Moreover, the code produces a number of diagnostic outputs
and allows for an iterative re-calibration of the results.
Figure \ref{method} shows a schematic depiction of the basic
PSF-matched fitting algorithm used in the code.

As HRI priors \textsc{t-phot} can use i) real cutouts of sources from the 
HRI, ii) models of sources obtained with \textsc{Galfit} or similar
codes, iii) a list of coordinates where PSF-shaped sources will be
placed, or a combination of these three types of priors.

For a detailed technical description of the mode of operation of the code,
we refer the reader to the Appendix and to the documentation included 
in the downloadable tarball. Here, we will briefly describe its main features.

\begin{figure*}[ht] 
\includegraphics[width=14cm]{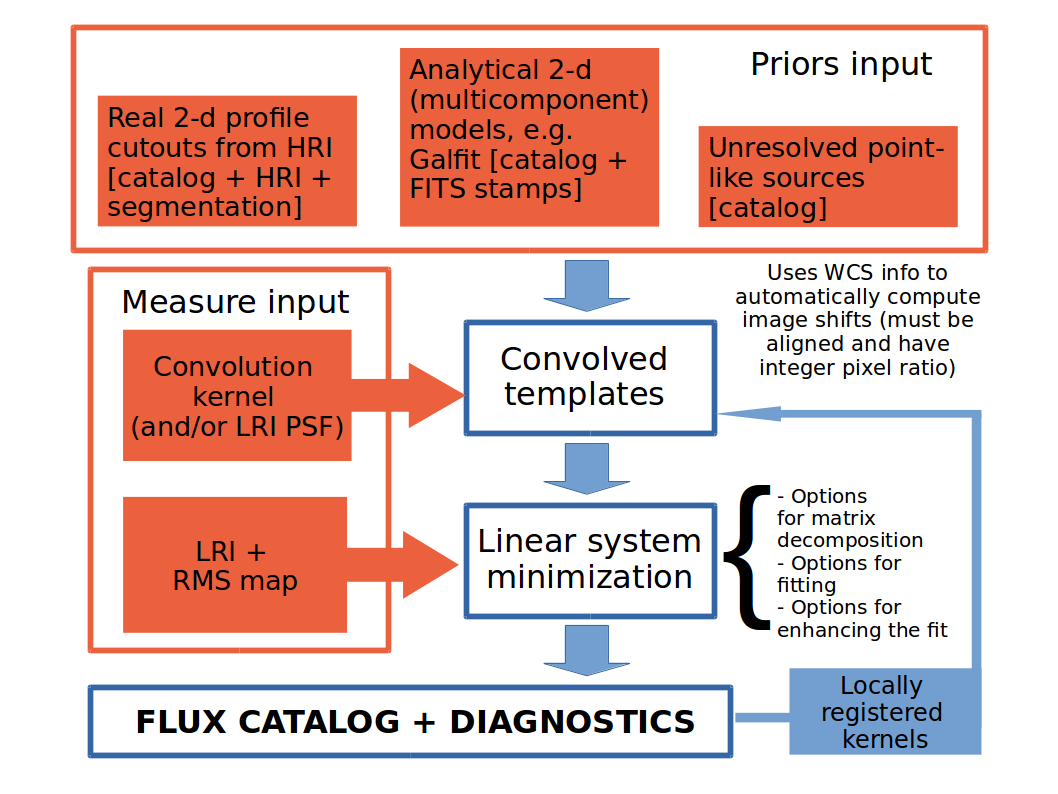}
\centering
\caption{Schematic representation of the workflow 
in \textsc{t-phot}.}\label{flow}
\end{figure*}

\subsection{Pipeline}

The pipeline followed by \textsc{t-phot} is outlined in the 
flowchart given in Fig. \ref{flow}. The following paragraphs 
give a short description of the pipeline.

\subsubsection{Input} \label{input}

The input files needed by \textsc{t-phot}
vary depending on the type(s) of priors used.

If true high-resolution priors are used, e.g. 
for optical/NIR ground-based or IRAC measurements using 
Hubble Space Telescope (HST) cutouts, \textsc{t-phot} needs
\begin{itemize}
\item the detection, high-resolution image (HRI) in \texttt{.fits} 
format;
\item the catalogue of the sources in the HRI, obtained using 
\textsc{SExtractor} or similar codes (the required format is
described in Appendix A);
\item the segmentation map of the HRI, in \texttt{.fits} format, 
again obtained using \textsc{SExtractor} or similar codes, 
having the value of the \texttt{id} of each source 
in the pixels belonging to it, and zero everywhere else;
\item a convolution kernel $K$, in the format of a \texttt{.fits} 
image or of a \texttt{.txt} file, matching the PSFs of the 
HRI and the LRI so that $PSF_{LRI}=K * PSF_{HRI}$ ($*$ 
is the symbol for convolution). The kernel must
have the HRI pixel scale.
\end{itemize}

If analytical models are used as priors (e.g. 
\textsc{Galfit} models), \textsc{t-phot} needs
\begin{itemize}
\item the stamps of the models (one per object, in \texttt{.fits} format);
\item the catalogue of the models (the required format is
described in Appendix A);
\item the convolution kernel $K$ matching the PSFs of
the HRI and the LRI, as in the previous case.
\end{itemize}
If models have more than one component, one separate 
stamp per component and catalogues for each component are needed 
(e.g. one catalogue for bulges and one catalogue for disks).

If unresolved, point-like priors are used, \textsc{t-phot} needs
\begin{itemize} 
\item the catalogue of positions (the required format is
described in Appendix A);
\item the LRI PSF, in the LRI pixel scale.
\end{itemize}

In this case, a potential limitation to the reliability of the method
is given by the fact that the prior density usually needs to
be optimized with respect to FIR/sub-mm maps, as discussed in 
\citet[][ in preparation]{Shu2015} and \citet{Elbaz2011} 
\citep[see also][ in preparation]{Wang2015, Bourne2015}.
The optimal number of priors turns out to be around
50-75\% of the numbers of beams in the map. The main problem is
identifying which of the many potential priors from, for example,
an HST catalogue one should use. This is a very complex issue
and we do not discuss it in this paper.

If mixed priors are used, \textsc{t-phot} obviously needs the
input files corresponding to each of the different types of priors
in use.

Finally, in all cases \textsc{t-phot} needs
\begin{itemize}
\item the measure LRI, background subtracted
(see next paragraph), in \texttt{.fits} format, with the same orientation 
as the HRI (i.e. no rotation allowed); the pixel scale can be equal to, 
or an integer multiple of, the HRI pixel scale, and the origin of one pixel
must coincide; it should be in surface brightness units (e.g. counts/s/pixel, 
or Jy/pixel for FIR images, and not PSF-filtered);
\item the LRI RMS map, in \texttt{.fits} format, with the 
same dimensions and WCS 
of the LRI.
\end{itemize}

Table \ref{tabinput} summarizes the input requirements for the different
choices of priors just described.

\begin{table*}[t!]
\begin{center}
  \begin{tabular}{ | l || c | c | c | }
    \hline
    & \textbf{Real cutouts} & \textbf{Analytical models} & \textbf{Point-sources} \\ \hline \hline
    \textbf{Priors} & \begin{tabular}{c} HRI \\ Segmentation \\ Catalogue \end{tabular} & \begin{tabular}{c} HRI \\ Model Stamps \\ Catalogue \end{tabular} & Positions Catalogue \\ \hline
    \textbf{Transformation} & Convolution Kernel & Convolution Kernel & PSF$_{LRI}$\\ \hline
    \textbf{Measure} & \begin{tabular}{c} LRI \\ RMS$_{LRI}$ \end{tabular} & \begin{tabular}{c} LRI \\ RMS$_{LRI}$ \end{tabular} & \begin{tabular}{c} LRI \\ RMS$_{LRI}$ \end{tabular} \\ \hline
  \end{tabular}
\end{center}
\caption{The input files needed by \textsc{t-phot} for
different settings. See text for details.} \label{tabinput}
\end{table*}

All the input images must have the following keywords in their headers:
\texttt{CRPIXn, CRVALn, CDn\_n, CTYPEn} (n=1,2).

\subsubsection{Background subtraction}

As already mentioned, the LRI must be background 
subtracted before being fed to \textsc{t-phot}. This
is of particular interest when dealing with FIR/sub-mm images,
where the typical standard is to use zero-mean.
To estimate the background level in optical/NIR images, 
one simple possibility is to take advantage of the 
option to fit point-like sources to measure the flux for
a list of positions chosen to fall within void regions.
The issue is more problematic in such confusion-limited FIR
images where there are no empty sky regions. 
In such cases, it is important to separate the fitted
sources (those listed in the prior catalogue) from the 
background sources, which contribute to a flat background level 
behind the sources of interest. The priors should be chosen so 
that these two populations are uncorrelated. The average
contribution of the faint background source population 
can then be estimated for example by i) injecting fake sources into 
the map and measuring the average offset
(output-input) flux, or ii) measuring the modal value 
in the residual image after a first pass through 
\textsc{t-phot} \citep[see e.g.][ in preparation]{Bourne2015}.

\subsubsection{Stages}

\textsc{t-phot} goes through ``stages'', each of which performs a 
well-defined task. The best results are obtained by performing two runs 
(``pass 1'' and ``pass 2''), the second using locally 
registered kernels produced during the first.
The possible stages are the following:

\begin{itemize}
\item \texttt{priors}: creates/organizes stamps for sources as listed 
in the input priors catalogue(s);
\item \texttt{convolve}: convolves each high-resolution stamp  
with the convolution kernel $K$ to obtain models (``templates'') 
of the sources at LRI resolution. The templates are normalized
to unit total flux. If the pixel scale of the images is 
different, transforms templates accordingly. 
Convolution is preferably performed in Fourier space,
using fast \texttt{FFTW3} libraries; however the user can
choose to perform it in real pixel space, ensuring
a more accurate result at the expense of a much 
slower computation. 
\item \texttt{positions}: if an input catalogue of 
unresolved sources is given, creates the PSF-shaped 
templates listed in it, and merges it with the one produced 
in the \texttt{convolve} stage;
\item \texttt{fit}: performs the fitting procedure, 
solving the linear system and obtaining the multiplicative
factors to match each template flux with the measured one;
\item \texttt{diags}: selects the best fits\footnote{Each source is fitted 
more than once if an arbitrary grid is used, as in the standard 
\textsc{tfit} approach.} and produces the final formatted output catalogues 
with fluxes and errors, plus some other diagnostics, see Sect. \ref{output};
\item \texttt{dance}: obtains local convolution kernels for the 
second pass; it can be skipped if the user is only
interested in a single-pass run;
\item \texttt{plotdance}: plots diagnostics for the dance stage; it
can be skipped for any purpose other than diagnostics;
\item \texttt{archive}: archives all results in a subdirectory whose name
is based on the LRI and the chosen fitting method (to be used only
at the end of the second pass).
\end{itemize}

The exact pipeline followed by the code is specified by a keyword
in the input parameter file. See also Appendix A for a more detailed
description of the whole procedure.

\begin{figure*}[ht]
\includegraphics[width=12cm]{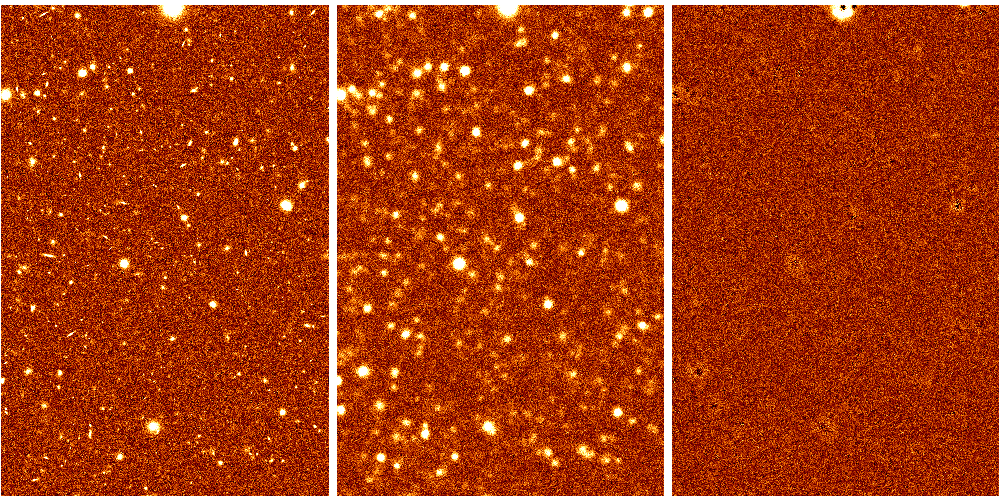}
\centering
\caption{Example of the results of a standard \textsc{t-phot} run
using extended priors. Left to right: HRI (FWHM=0.2"), 
LRI (FWHM=1.66"), and residuals image for a simulated dataset.
LRI and residual image are on the same greyscale.}\label{example}
\end{figure*}

\begin{figure*}[ht] 
\includegraphics[width=11cm]{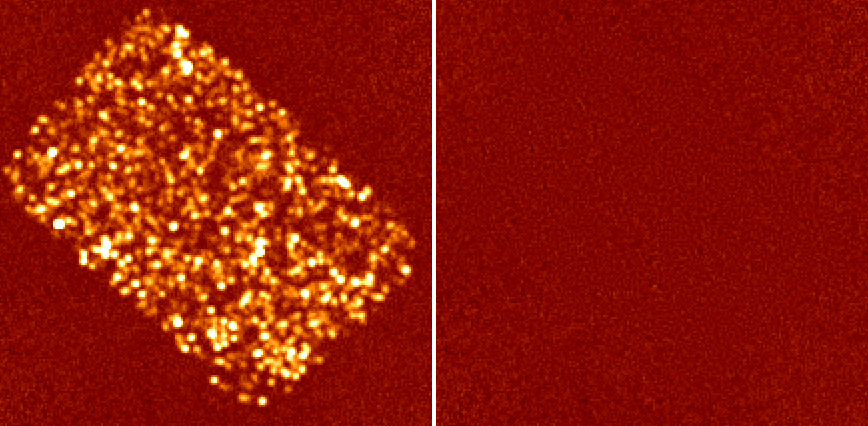}
\centering
\caption{Example of the results of a standard \textsc{t-phot} run 
using point-source priors. Left to right: LRI (FWHM=25'') 
and residuals image (same greyscale) for a simulated dataset.
See also Sect. \ref{realistic}.}\label{herschel}
\end{figure*}

\begin{figure*}[ht] 
\includegraphics[width=14cm]{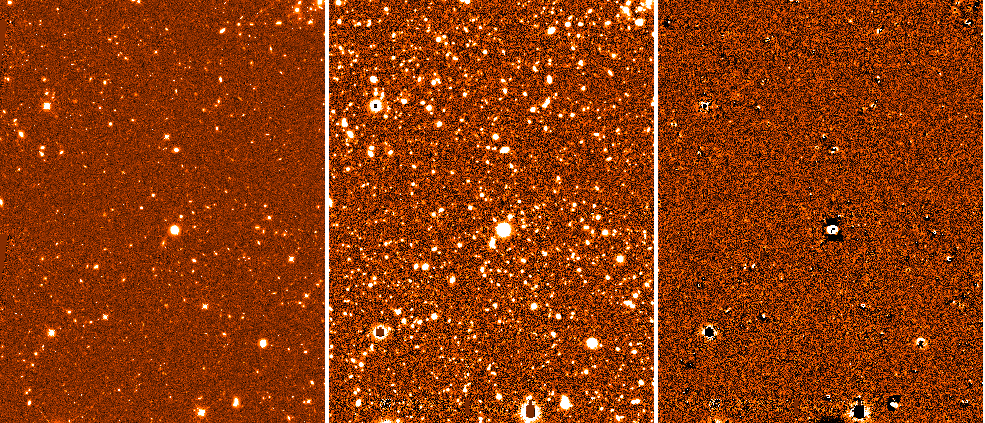}
\centering
\caption{Example of the results of a standard \textsc{t-phot} run 
using analytical priors. Left to right: CANDELS COSMOS $H$-band (HRI), 
$R$-band (LRI) and residuals image obtained 
using \textsc{Galfit} two-component models. 
LRI and residual image are on the same greyscale.}\label{cosmos}
\end{figure*}

\subsubsection{Solution of the linear system} \label{linsys}

The search for the LRI fluxes of the objects detected in the HRI
is performed by creating a linear system
\begin{equation}
\sum_{m,n}I(m,n) = \sum_{m,n} \sum_i^N F_iP_i(m,n)
\end{equation}
\noindent where $m$ and $n$ are the pixel indexes, 
$I$ contains the pixel values of the fluxes in the LRI, 
$P_i$ is the normalized flux of the template for the $i$-th objects in 
the (region of the) LRI being fitted, and $F_i$ is the multiplicative scaling 
factor for each object. In physical terms, $F_i$ represents the flux of 
each object in the LRI (i.e. it is the unknown to be determined).

Once the normalized templates for each object in the LRI 
(or region of interest within the LRI) have been generated during the 
\texttt{convolve} stage, the best 
fit to their fluxes can be simultaneously derived by minimizing a 
$\chi^2$ statistic, 
\begin{equation}
\chi^2=\left[ \frac{\sum_{m,n}I(m,n)-M(m,n)}{\sigma(m,n)} \right]^2
\end{equation}

\noindent where $m$ and $n$ are the pixel indexes,
\begin{equation}
M(m,n)=\sum_i^N F_iP_i(m,n)
\end{equation}

\noindent and $\sigma$ is the value of the RMS map
at the $(m,n)$ pixel position.

The output quantities are the best-fit solutions of the minimization procedure, 
i.e. the $F_i$ parameters and their relative errors. 
They can be obtained by resolving the linear system
\begin{equation}
\frac{\partial \chi^2}{\partial F_i} = 0
\end{equation}

\noindent for $i=0,1,...,N$.

In practice, the linear system can be rearranged into a matrix equation,
\begin{equation}
AF=B
\end{equation}

\noindent where the matrix $A$ contains the coefficients $P_iP_j/\sigma^2$, 
$F$ is a vector containing the fluxes to be determined, and $B$ is a vector
given by $I_iP_i/\sigma^2$ terms. The matrix equation is solved via
one of three possible methods as described in the next subsection.

\subsubsection{Fitting options} \label{news}
\textsc{t-phot} allows for some different options to perform
the fit: 

\begin{itemize}
\item three different methods for solving the linear system are 
implemented, namely, the LU method (used by default in \textsc{tfit}), 
the Cholesky method, and the Iterative Biconjugate Gradient method 
\citep[used by default in \textsc{convphot}; 
for a review on methods to solve sparse
linear systems see e.g.][]{Davis2006}. They yield similar
results, although the LU method is slightly more stable and faster;
\item a threshold can be imposed so that only pixels with a flux
higher than this level will be used in the fitting procedure 
(see Sect. \ref{threshold});
\item sources fitted with a large, unphysical negative flux
($f_{meas}<-3\sigma$, where $\sigma$ is their nominal error, see below) 
can be excluded from the fit, and in this case 
a new fitting loop will be performed without considering these sources.
\end{itemize}

The fit can be performed i) on the entire LRI as a whole,
producing a single matrix containing all the sources (this is
the method adopted in \textsc{convphot}); 
ii) subdividing the LRI into an arbitrary grid of (overlapping) small 
cells, perfoming the fit in each of such cells separately, and then 
choosing the best fit for each source, using some convenient criteria 
to select it (because sources will be fitted more than once if the cells 
overlap; this is the method adopted in \textsc{tfit}); iii) ordering 
objects by decreasing flux, building a cell around each source 
including all its potential contaminants, solving the problem 
in that cell and assigning to the source the obtained flux 
(\textit{cells-on-objects} method; see the Appendix for more details).

While the first method is the safest and more accurate because
it does not introduce any bias or arbitrary modifications, it may often
be unfeasible to process at once large or very crowded images.
Potentially large computational time saving is possible using the 
\textit{cells-on-objects} method, depending on the level of 
blending/confusion in the LRI; if it is very high, most sources
will overlap and the cells will end up being very large.
This ultimately results in repeating many times the fit on regions with
dimensions comparable to the whole image (a check is implemented
in the code, to automatically change the method from
\textit{cells-on-objects} to single fit if this is the case). If the 
confusion is not dramatic, a saving in computational time up to two
orders of magnitude can be achieved. The results obtained using
the \textit{cells-on-objects} method prove to be virtually
identical to those obtained with a single fit on the whole image 
(see Sect. \ref{realistic}).
On the other hand, using the arbitrary cells method 
is normally the fastest option, but can introduce potentially
large errors to the flux estimates owing to wrong assignments of
peripheral flux from sources located outside a given cell to 
sources within the cell (again, see Sect. \ref{realistic} 
and the Appendix B). 

\subsubsection{Post-fitting stages: kernel registration}

After the fitting procedure is completed, \textsc{t-phot} will produce the
final output catalogues and diagnostic images (see Sect. \ref{output}). Among these,
a \textit{model} image is obtained by adding all the templates, 
scaled to their correct total flux after fitting, in the positions
of the sources. This image will subsequently be used if a second pass 
is planned; during a stage named \texttt{dance}, 
a list of positional shifts is computed, and a set of shifted kernels 
are generated and stored. The \texttt{dance} stage consists of 
three conceptual steps:
\begin{itemize}
\item the LRI is divided into cells of a given size (specified 
by the keyword \texttt{dzonesize}) and a linear $\Delta x, \Delta y$ 
shift is computed within each cell, cross-correlating the model image 
and the LRI in the considered region\footnote{FFT and direct 
cross-correlations are implemented, the latter being
the preferred default choice because it gives more precise 
results at the expense of a slightly slower computation.};
\item interpolated shifts are computed for the regions where
the previous registration process gives spuriously large
shifts, i.e. above the given input threshold parameter
\texttt{maxshift};
\item the new set of kernels is created using the computed shifts
to linearly interpolate their positions, while catalogues 
reporting the shifts and the paths to kernels are produced.
\end{itemize}

\subsubsection{Second pass}

The registered kernels can subsequently be used in the second 
pass run to obtain more astrometrically precise results. \textsc{t-phot}
automatically deals with them provided the correct keyword is given in the
parameter file. If unresolved priors are used, the list of shifts generated
in the \texttt{dance} stage will be used by the \texttt{positions} 
routine during the second pass to produce correctly shifted PSFs
and generate new templates.

\subsection{Error budget} \label{errors}

During the fitting stage, the covariance matrix is constructed. 
Errors for each source are assigned as the 
square root of the diagonal element of the 
covariance matrix relative to that source. It must be pointed out that using 
any cell method for the fitting rather than the single fitting option 
will affect this uncertainty budget, since a different matrix will be 
constructed and resolved in each cell. 

It is important to stress that this covariance error budget
is a \emph{statistical} uncertainty, relative to the RMS fluctuations
in the measurement image, and is not related to any possible 
\emph{systematic} error. The latter can instead be estimated by
flagging potentially problematic sources, to be identified
separately from the fitting procedure.
There can be different possible causes for systematic
offsets of the measured flux with respect to the true
flux of a source. \textsc{t-phot} assigns the following flags:
\begin{itemize}
\item +1 if the prior has saturated or negative flux;
\item +2 if the prior is blended (the check
is performed on the segmentation map);
\item +4 if the source is at the border of the image (i.e. its
segmentation reaches the limits of the HRI pixels range).
\end{itemize}

\subsection{Description of the output} \label{output}

\textsc{t-phot} output files are designed to be very similar in format
to those produced by \textsc{tfit}. They provide
\begin{itemize}
\item a ``best'' catalogue containing the following data, listed for each
detected source (as reported in the catalogue file header): 
\begin{itemize}
\item \texttt{id}; 
\item \texttt{x} and \texttt{y} positions (in LRI pixel scale 
and reference frame, FITS convention where the first pixel is 
centred at 1,1);
\item \texttt{id} of the cell in which the best fit 
has been obtained (only relevant for the arbitrary grid fitting method);
\item $x$ and $y$ positions of the object in the cell and distance
from the centre (always equal to 0 if the 
\textit{cells-on-objects} method is adopted);
\item fitted flux and its uncertainty (square root of
the variance, from the covariance matrix). 
These are the most important output quantities;
\item flux of the object as given in the input HRI catalogue or,
in the case of point-source priors, measured flux of the 
pixel at the $x,y$ position of the source in the LRI; 
\item flux of the object as determined in the \texttt{cutout} 
stage (it can be different to the previous one, 
e.g. if the segmentation was dilated); in the case of
point-sources priors, measured flux of the 
pixel at the $x,y$ position of the source in the LRI;
\item \emph{flag} indicating a possible bad 
source as described in the previous subsection; 
\item number of fits for the object (only relevant for arbitrary 
grid fitting method, 1 in all other cases). 
\item \texttt{id} of the object having the largest covariance 
with the present source; 
\item \textit{covariance index}, i.e. the ratio of the maximum
covariance to the variance of the object itself;
this number can be considered an indicator of the
reliability of the fit, since large covariances
often indicate a possible systematic offset in the
measured flux of the covarying objects (see Sect. \ref{realistic}).
\end{itemize}
\item two catalogues reporting statistics for the fitting cells and the 
covariance matrices (they are described in the documentation);
\item the model \texttt{.fits} image, obtained as a collage of the templates,
as already described;
\item a diagnostic residual \texttt{.fits} image, obtained by subtracting the 
model image from the LRI;
\item a subdirectory containing all the low-resolution model templates;
\item a subdirectory containing the covariance matrices in graphic 
(\texttt{.fits}) format;
\item a few ancillary files relating to the shifts of the kernel
for the second pass and a subdirectory containing the shifted kernels.
\end{itemize}

All fluxes and errors are output in units consistent with the
input images.

Figures \ref{example}, \ref{herschel} and \ref{cosmos} show
three examples of \textsc{t-phot} applications on simulated
and real data, using the three different options for
priors.

\section{Assumptions and limitations} \label{limits}

\begin{figure}[ht] 
\includegraphics[width=4.5cm]{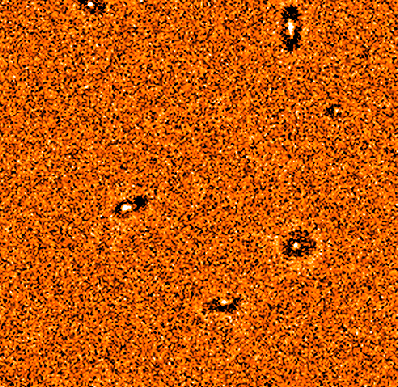}
\centering
\caption{Typical patterns in a residual image created
by \textsc{t-phot}, caused by inaccurate PSF/kernel 
determination. In this case, the ring-shaped shadow
surrounding a bright central spot is due to an
underestimation of the central peak of the LRI PSF,
which causes an overestimation of the fit in the 
outskirts while leaving too much light in the centre.}\label{PSF}
\end{figure}

The PSF-matching algorithms implemented in
\textsc{t-phot} and described in the previous section
are prone to some assumptions and limitations. In particular,
the following issues must be pointed out.

i) The accuracy of the results strongly depends on the 
reliability of the determined PSFs 
(and consequently of the convolution kernel). 
An error of a few percentage points in the central slope of the 
PSF light profile might lead to non-negligible systematical 
deviations in the measured fluxes. 
However, since the fitting algorithm minimizes the residuals 
on the basis of a summation over pixels, an incorrect PSF 
profile will lead to characteristic positive and
negative ring-shaped patterns in the residuals (see 
Fig. \ref{PSF}), and to some extent the summation over 
pixels will compensate the global flux determination.

ii) When dealing with extended priors, it is assumed 
that the instrinsic morphology of the objects does not 
change with the wavelength. Of course, this is usually 
not the case. The issue is less of a problem when dealing with 
FIR images, in which the morphological features of the priors 
are unresolved by the low-resolution PSF.
On the other hand, in the optical and NIR domains this 
problem may be solved by the use of multicomponent 
analytical models as priors. In this approach, each component should
be fitted independently, thus allowing the ratio between 
bulge and disk components to vary between the HRI and LRI.  
A clear drawback of this approach is that any failure of the fit 
due to irregular or difficult morphological features 
(spiral arms, blobs, asymmetries, etc.) would be propagated 
into the LRI solution. This functionality is already implemented 
in \textsc{t-phot} and detailed testing is ongoing.

iii) As explained in Sect. \ref{errors}, \textsc{t-phot} flags 
priors that are likely to be flawed: sources too close to 
the borders of the image, saturated objects, and most notably
\emph{blended priors}. The assumption that all priors are
well separated from one another is crucial, and the method 
fails when this requirement is not accomplished. 
Again, this is crucial only when dealing
with real priors, while analytical models and unresolved 
priors are not affected by this limitation.

iv) As anticipated in Sect. \ref{input}, FIR images can suffer
from an ``overfitting'' problem, due to the presence of too many 
priors in each LRI beam if the HRI is deeper than the LRI. In this case,
a selection of the priors based on some additional criteria (e.g.
flux predition from SED fitting) might be necessary to avoid catastrophic 
outcomes \citep[see also][ in preparation]{Wang2015, Bourne2015}.

\begin{figure*}[ht] 
\includegraphics[width=7cm,height=4.5cm]{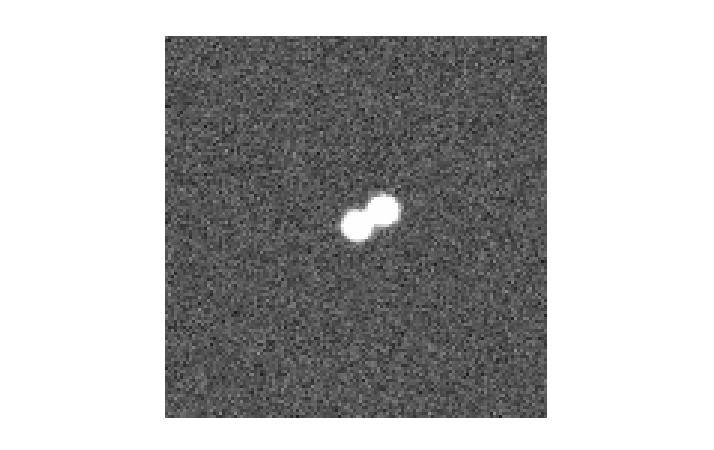}
\includegraphics[width=10cm,height=6cm]{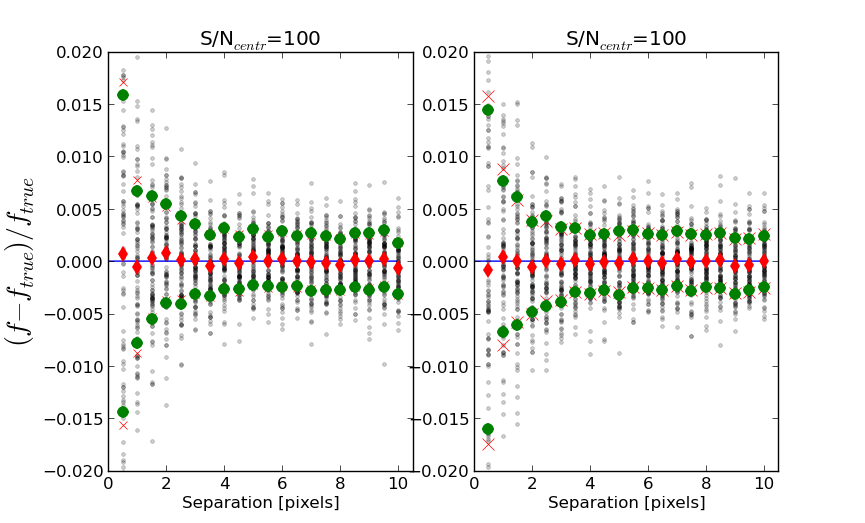}
\includegraphics[width=7cm,height=4.5cm]{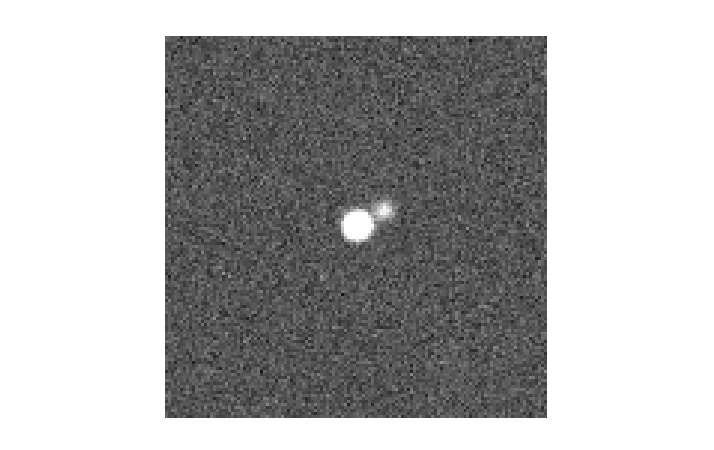}
\includegraphics[width=10cm,height=6cm]{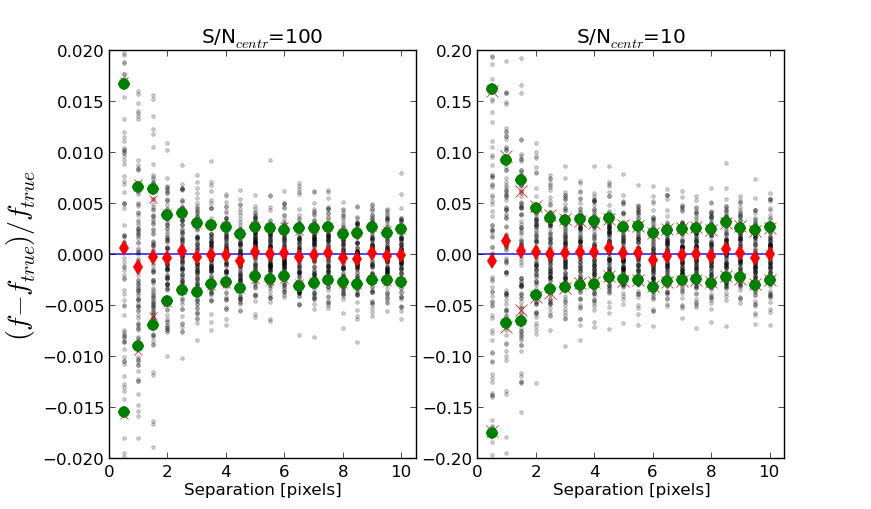}
\includegraphics[width=7cm,height=4.5cm]{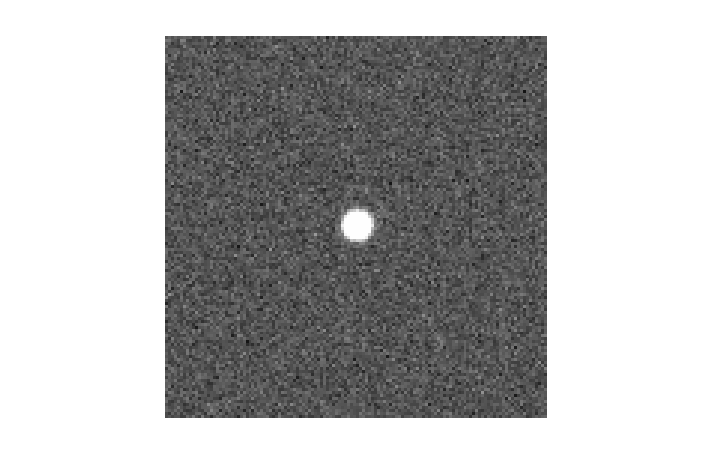}
\includegraphics[width=10cm,height=6cm]{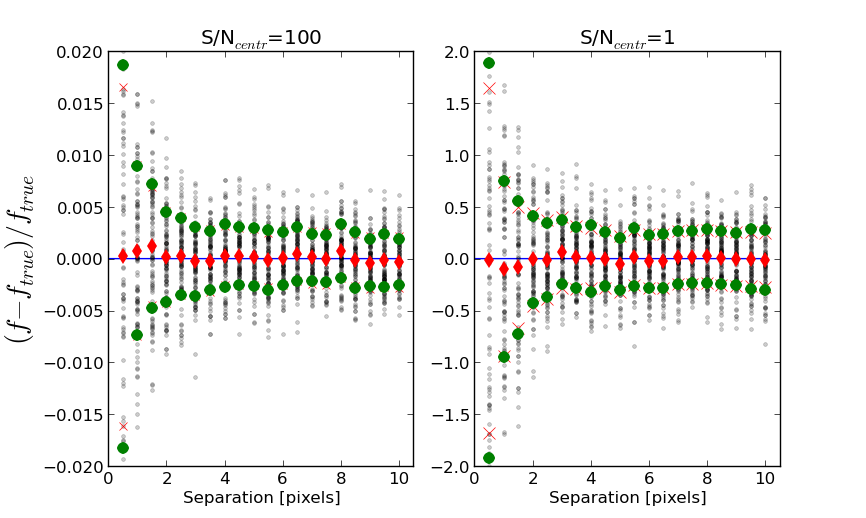}
\centering
\caption{Accuracy check on idealized PSF-shaped objects. 100 realizations 
of the same image containing two PSF-shaped objects at varying positions 
and signal-to-noise ratios have been produced and the fluxes have been 
measured with \textsc{t-phot}.
In each row, the left image shows one of the 100 realizations with the largest
considered separation (10 pixels). On the right, the first panel refers to 
the central object, and the second (on the right) to the shifted object; 
central signal-to-noise ($S/N_{centr}$) ratios are, from top to bottom, 
100, 100, 100 for the first source and 100, 10, 1 for the second source.
In each panel, as a function of the separation interval between 
the two sources, the faint grey points show each of the 100 
flux measurements (in relative difference with respect 
to the true input flux), the red diamonds are the averages of the
100 measurements, the red crosses show the nominal error given by the 
covariance matrix in \textsc{t-phot}, and the green dots
the standard deviation of the 100 measurements. See text for more 
details.}\label{koryo}
\end{figure*}

\begin{figure*}[ht]
\includegraphics[width=12cm]{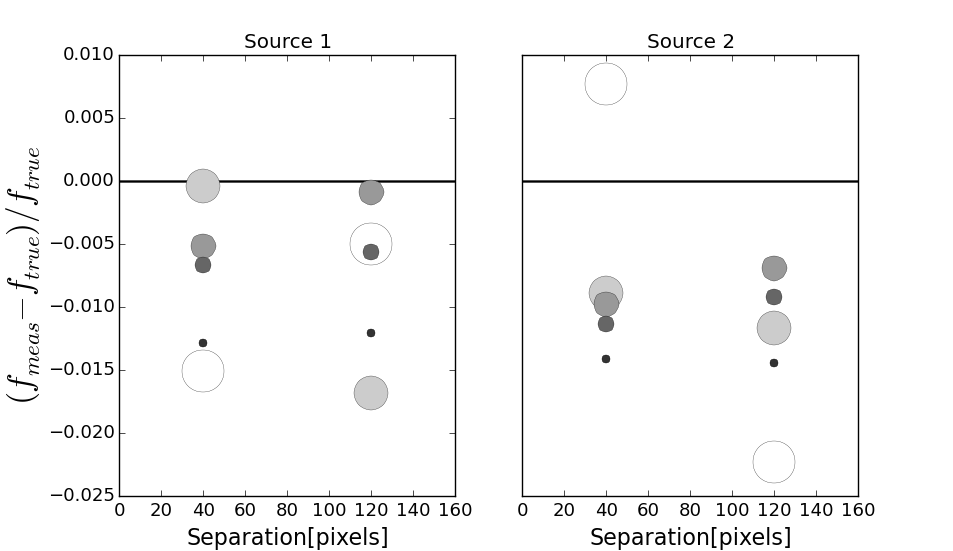}
\centering
\caption{Effects of different segmentation areas
on the measured flux of two isolated 
objects with identical flux and signal-to-noise ratio,
at two possible separations of 40 and 120 pixels. 
Each panel shows the flux error in one of the objects 
at each separation distance.
The shades and dimensions of the dots is a function 
of the radius of the segmentation, with darker and smaller 
dots corresponding to smaller segmentations.
See text for more details.}\label{segmentation}
\end{figure*}

\begin{figure*}[ht] 
\includegraphics[width=14cm]{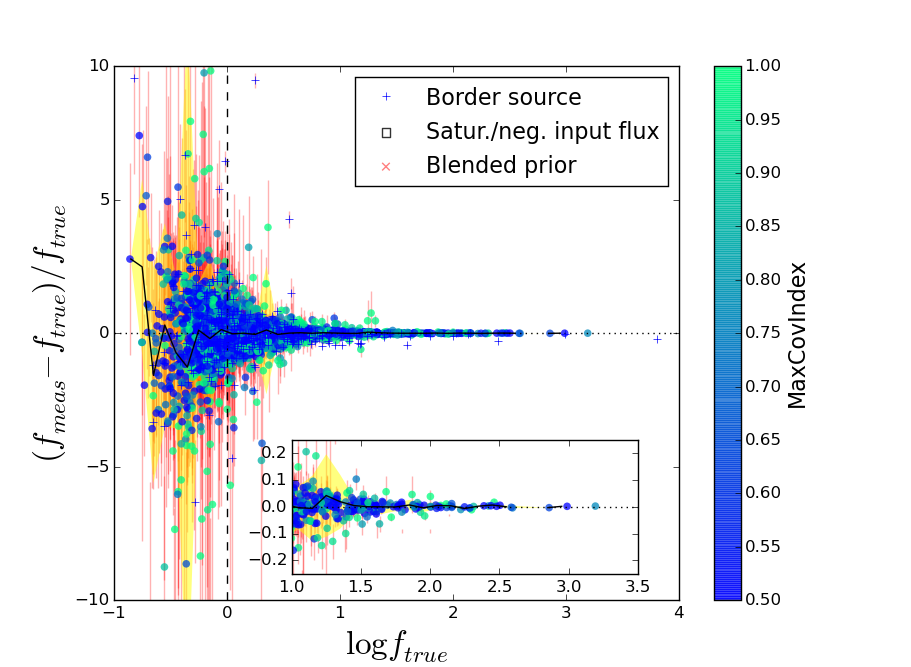}
\centering
\caption{Accuracy of the flux determination in a simulation
containing non-overlapping, PSF-shaped sources and 
``perfect'' detection. Relative measured flux
difference $(f_{meas}-f_{true})/f_{true}$ is plotted versus logarithm of
the input flux $f_{true}$, for a simulated image populated with PSF-shaped 
sources (FWHM=1.66"). Each dot corresponds to a single
source, with different symbols and colours referring to various 
diagnostics as explained in the legend and in the colourbar.
The black solid line is the average in bins, the yellow shade is 
the standard deviation. The vertical dashed line shows the limiting flux
at 1$\sigma$, $f=1$. The inner panel shows a magnification
of the brighter end of the distribution.
The fit was performed on the whole 
image at once. See text for more details.}\label{accuracy1}
\end{figure*}

\begin{figure*}[ht] 
\includegraphics[width=8cm]{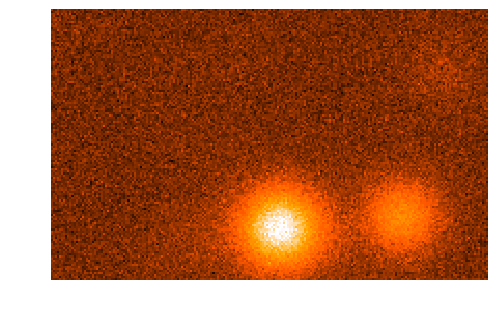} 
\includegraphics[width=9.5cm]{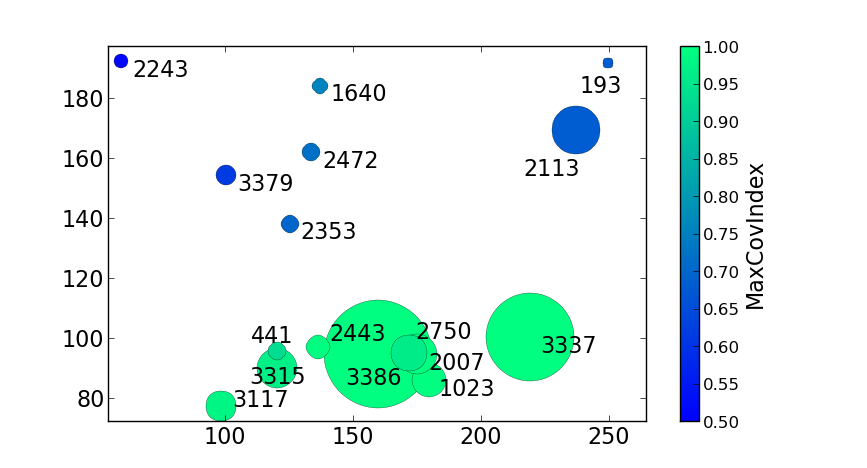}
\includegraphics[width=9.5cm]{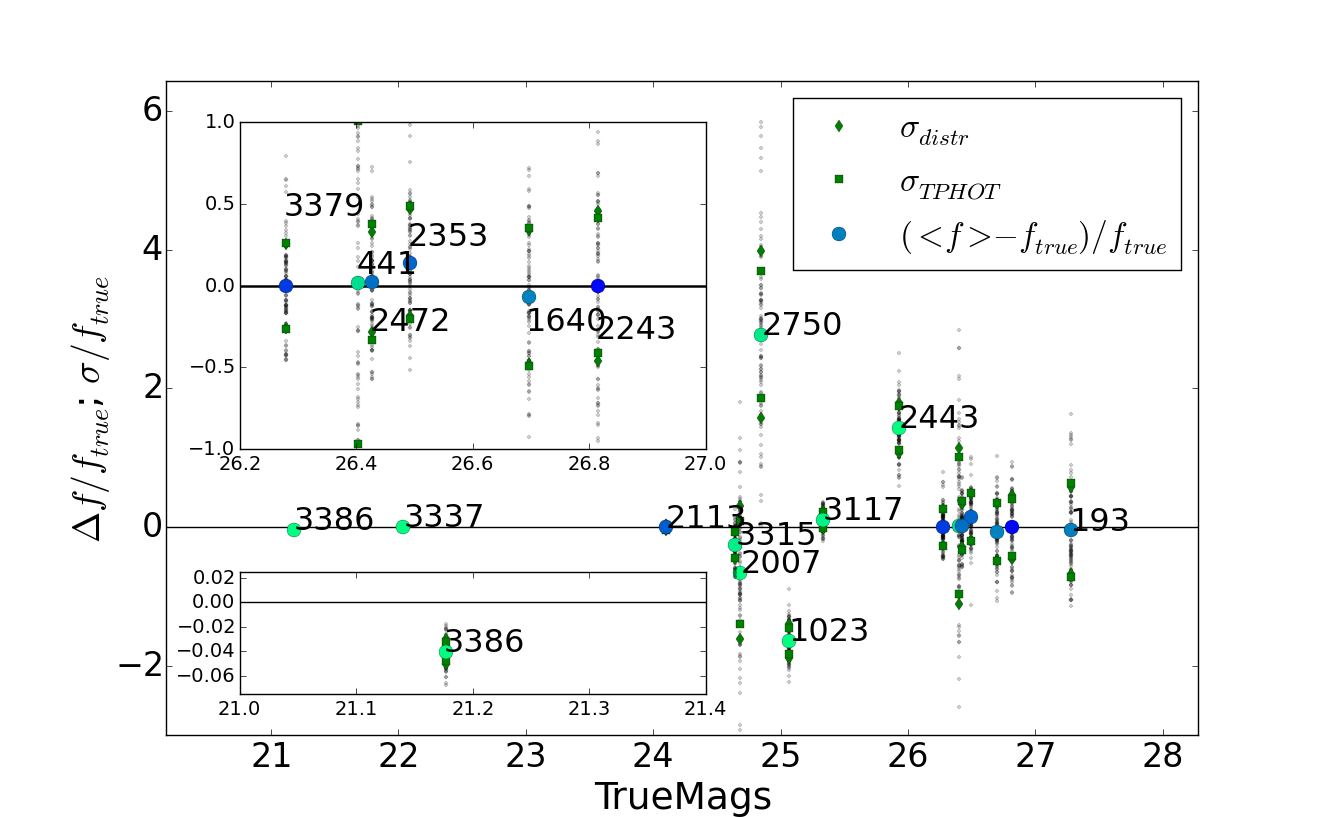}
\includegraphics[width=8cm]{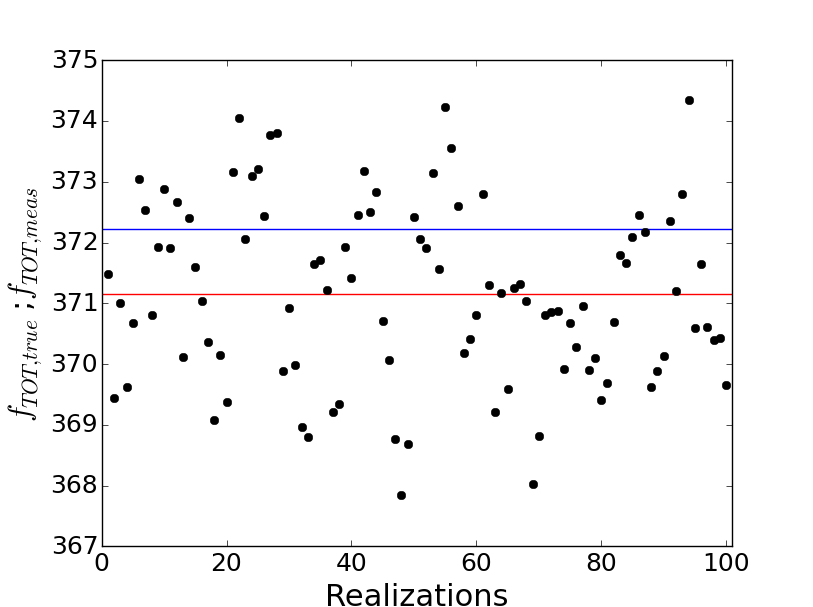}
\centering
\caption{Analysis of a small region including a strongly covarying 
group of sources. Upper left panel: one of the 100 realizations with different
noise maps of the region. Upper right panel: true spatial position of all 
the sources in the region (the colour of the dots refer to the covariance 
index of the sources, as indicated in the colourbar, while their size is 
proportional to their true flux). Bottom left panel: relative deviation of
measured flux from the true flux for each source in the region,
as a function of their true magnitude (big dots show the average 
relative deviation, and their colours refer to their covariance
index as in the previous panel; green squares show the nominal uncertainty
given by \textsc{t-phot}, to be compared with the RMS of the distribution
of the 100 measurements (diamonds); small grey dots are 
the 100 measurements. The insets show magnifications of regions of
interest). Bottom right panel: each dot shows the sum of the measured fluxes 
for each of the 100 realizations, and the average of this sum (red line)
to be compared with the true sum (blue line), showing that
an overall consistency is guaranteed by the method. 
See text for more details.}\label{region}
\end{figure*}

\section{Validation} \label{validation}

To assess the performance of \textsc{t-phot} we set up an extensive set
of simulations, aimed at various different and complementary goals.

We used \textsc{SkyMaker} \citep{Bertin2009}, 
a public software tool, to build synthetic \texttt{.fits} images. The
code ensures direct control on all the observational parameters 
(the magnitude and positions of the objects, their morphology,
the zero point magnitude, the noise level, and the PSF). 
Model galaxies were built by summing a de Vaucouleurs and an exponential 
light profile in order to best mimic a realistic distribution 
of galaxy morphologies. These models were generated using a variety 
of bulge-to-total light ratios, component sizes, and projection angles. 

All tests were run using ideal (i.e. synthetic and symmetric) 
PSFs and kernels.

Moreover, we also performed tests on real datasets taken from the CANDELS survey
(in these cases using real PSFs).

Some of the tests were performed using both \textsc{t-phot}
and \textsc{tfit}, to cross-check the results, ensuring
the perfect equivalency of the results given by the two
codes when used with the same parameter sets, and showing how
appropriate settings of the \textsc{t-phot} parameters can ensure
remarkable improvements.

For simplicity, here we only show the results 
from a restricted selection of the test dataset, 
which are representative of the performance of \textsc{t-phot}
in standard situations. The results of the other simulations
resemble overall the ones we present, and are omitted for the
sake of conciseness.

\subsection{Code performance and reliability on simulated images}

\subsubsection{Basic tests}

As a first test, we checked the performance of the basic method
by measuring the flux of two PSF-shaped synthetic sources,
with varying separation and signal-to-noise ratios. One hundred
realizations with different noise maps of each parameter set were prepared, 
and the averages on the measured fluxes were computed. The aims of 
this test were twofold: on the one hand, to check the precision to which 
the fitting method can retrieve true fluxes in the simplest possible 
case - two sources with ideal PSF shape; on the other hand,
to check the reliability of the nominal error budget given by the 
covariance matrix, comparing it to the real RMS of the 100 measurements.
Figure \ref{koryo} shows three examples of the set-up and the results of 
this test. Clearly, in both aspects the results are reassuring: 
the average of the 100 measurements (red diamonds) is always in very good 
agreement with the true value, with offset in relative error always well 
under the $1/(S/N)_{centr}$ limit ($(S/N)_{centr}$ is the value of the signal-to-noise 
ratio in the central pixel of the source, corresponding to roughly 
one third of the total $S/N$); and the nominal error (red crosses) 
given by the covariance matrix is always in good agreement with the RMS 
of the 100 measurements (red circles).

When dealing with extended objects rather than with point-like
sources, one must consider the additional problem 
that the entire profile of the source cannot be measured exactly
because the segmentation is limited by the lowest signal-to-noise
isophote. The extension of the segmentation therefore 
plays a crucial role and defining it correctly is a very subtle problem.
Simply taking the isophotal area as reported by \textsc{SExtractor} as 
\texttt{ISOAREA} often underestimates the real extension of the objects. 
Accordingly, the segmentation of the sources should somehow be enlarged 
to include the faint wings of sources. To this aim, specific software 
called \textsc{Dilate} was developed at OAR and used in the CANDELS 
pipeline for the photometric analysis of GOODS-S and UDS IRAC data
\citep{Galametz2013}. \textsc{Dilate} enlarges the 
segmentation by a given factor, depending on the original area;
it has proven to be reasonably robust in minimizing the effects 
of underestimated segmentated areas.

Figure \ref{segmentation} shows the effects of artificially varying 
the dimensions of the segmentation relative to two bright, extended 
and isolated sources in a simulated HRI, on the flux measured
for that source in a companion simulated LRI.
It is important to note how enlarging the segmented area normally
results in larger measured fluxes, 
because more and more light from the faint wings of the source 
are included in the fit. However, beyond a certain limit 
the measurements begin to lose accuracy owing to the inclusion 
of noisy, too low signal-to-noise regions (which may cause 
a lower flux measurement). 

In principle, using extended analytical 
models rather than real high-resolution cutouts
should cure this problem more efficiently, because models have
extended wings that are not signal-to-noise limited. 
Tests are ongoing to check the performance of this 
approach, and will be presented in a forthcoming paper.

\subsubsection{Tests on realistic simulations} \label{realistic}

The next tests were aimed at investigating less idealized 
situations, and have been designed to provide a robust 
analysis of the performance of the code on realistic datasets. 
We used the code \textsc{GenCat} 
\citep[][ in preparation]{Schreiber2015} to produce mock 
catalogues of synthetic extragalactic sources, with reasonable 
morphological features and flux distribution\footnote{\textsc{GenCat} 
is another software package developed within the 
\textsc{astrodeep} project. It uses GOODS-S CANDELS statistics 
to generate a realistic distribution of masses at 
all redshifts, for two populations of galaxies (active and 
passive), consistently with observed mass functions. 
All the other physical properties of the mock galaxies are 
then estimated using analytical recipes from literature: 
each source is assigned a morphology (bulge-to-total ratio,
disk and bulge scale lengths, inclination etc.), 
star formation rate, attenuation, optical and infrared
rest-frame, and observed magnitudes. Each source is finally assigned 
a sky-projected position mimicking the clustering 
properties of the real CANDELS data.}.
Then, a set of images were produced using such catalogues as an input
for \textsc{SkyMaker}. A ``detection'' HRI mimicking an HST H 
band observation (FWHM = $0.2''$) was generated from the 
\textsc{GenCat} catalogue using output parameters to characterize 
the objects' extended properties. Then a set of measure LRIs 
were produced: the first was populated with PSF-shaped sources, 
having FWHM = 1.66'' (the typical IRAC-ch1/ch2 resolution,
a key application for \textsc{t-phot}), 
while other LRIs were created from the input 
catalogue, mimicking different ground-based or IRAC FWHMs.
Finally, we created another HRI catalogue removing all of the 
overlapping sources\footnote{We proceeded as follows. First,
we created a ``true'' segmentation image using the input catalogue 
and assigning to each object all the pixels in which the flux 
was $1.005 \times f_{background}$. Then, starting from 
the beginning of the list, we included each source in the 
new catalogue if its segmented area did not overlap 
the segmented area of another already inserted source.}. 
This ``non-overlapping'' catalogue was used to create parallel 
detection and measurement images in order to obtain insight into the 
complications given by the presence of overlapping priors. 
In all these images, the limiting magnitude was set equal to the
assigned zero point, so that the limiting flux at 1$\sigma$ is 1. 
In addition, the fits were always performed on the LRI as a whole, 
if not otherwise specified.

Figure \ref{accuracy1} shows the results relative to the first test,
i.e. the fit on the image containing non-overlapping, PSF-shaped 
sources, with a ``perfect'' detection (i.e. the priors catalogue
contains all sources above the detection limit),
obtained with a single fit on the whole image. The figure shows the
relative error in the measured flux of the sources, 
$(f_{meas}-f_{true})/f_{true}$, versus the log of the real 
input flux $f_{true}$; the different symbols
refer to the flag assigned to each object, while the colour
is a proxy for the covariance index. 

In this case, the only source of uncertainty in the measurement is given
by the noise fluctuations, which clearly become dominant at the
faint end of the distribution. Looking at the error bars of the
sources, which are given by the nominal error assigned by \textsc{t-phot}
from the covariance matrix, one can see that almost all sources
have measured flux within $2 \sigma$ from their true flux, 
with only strongly covariant sources (covariance index $\simeq$ 1, 
greener colours) having $|f_{meas}-f_{true}|/f_{true} > 1 \sigma$. 
The only noticeable exceptions are sources that have been flagged
as potentially unreliable, as described in Sect. \ref{errors}. 
We also note how the average $\Delta f/f$ (solid black line) 
is consistent with zero down to $f_{true} = S/N \simeq 0.63$.

Figure \ref{region} shows the analysis of a case study
in which the fluxes of a clump of highly covariant objects 
are measured with poor accuracy, and some of the nominal uncertainties
are underestimated: a very bright source (ID 3386, m$_{true}=21.17$) 
shows a relative difference $(f_{meas}-f_{true})/f_{true}>3 \sigma$. 
To cast light on the reason for such a discrepancy, the region 
surrounding the object was replicated 100 times with different 
noise realizations, and the results were analysed and compared. 
The upper panels show (left) one of the 100 measurement images 
and (right) the position of all the sources in the region 
(many of which are close to the detection limit). The colour 
code refers to the covariance index of the sources.
The bottom left panel shows the relative error in the measured 
flux for all the sources in the region, with the inner panels 
showing magnifications relative to the object ID 3386 and to 
the bunch of objects with m$_{true} \sim 26.5$.
Looking at the colours of their symbols, many objects in 
the region turn out to be strongly covariant. Indeed,
while the bluer sources in the upper part of the 
region all have covariant indexes lower than 0.5, the greener
ones in the crowded lower part all have covariance index
larger than 1 (indeed larger than 2 in many cases).
This means that their flux measurements are subject to 
uncertainties not only from noise fluctuations, but also from
systematic errors due to their extremely close and bright neighbours. 
As clearly demonstrated here, the covariance index can give a clue 
to which measurements can be safely trusted.

The bottom right panel gives the sum of the measured fluxes of 
all sources in each of the 100 realizations (the blue line 
is the true total flux and the red line is the mean of the 
100 measured total fluxes). It can be seen that the total
flux measured in the region is always consistent with the expected 
true one to within $\simeq 1\%$ of its value.

Although it is not possible to postulate a one-to-one relation 
(because in most cases sources having a large covariance index
have a relatively good flux estimate, see Fig. \ref{accuracy1}), 
the bottom line of this analysis is that 
the covariance index, together with the flagging code outputted by 
\textsc{t-phot}, can give clues about the reliability of measured flux,
and should be taken into consideration during the analysis of
the data. Measurements relative to sources having a covariance 
index larger than 1 should be treated with caution.

\begin{figure*}[ht] 
\includegraphics[width=14cm]{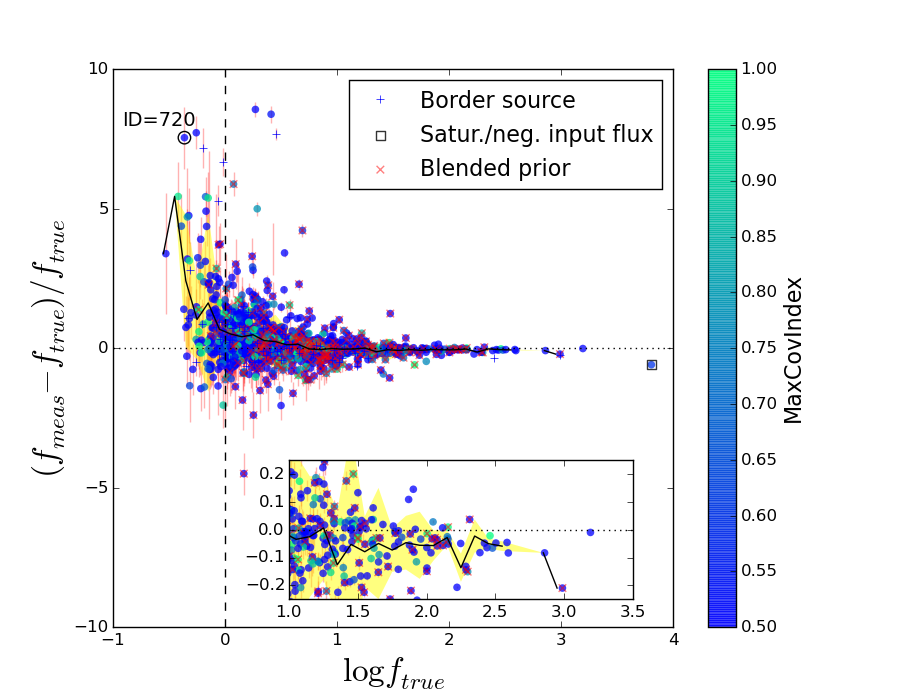}
\includegraphics[width=14cm]{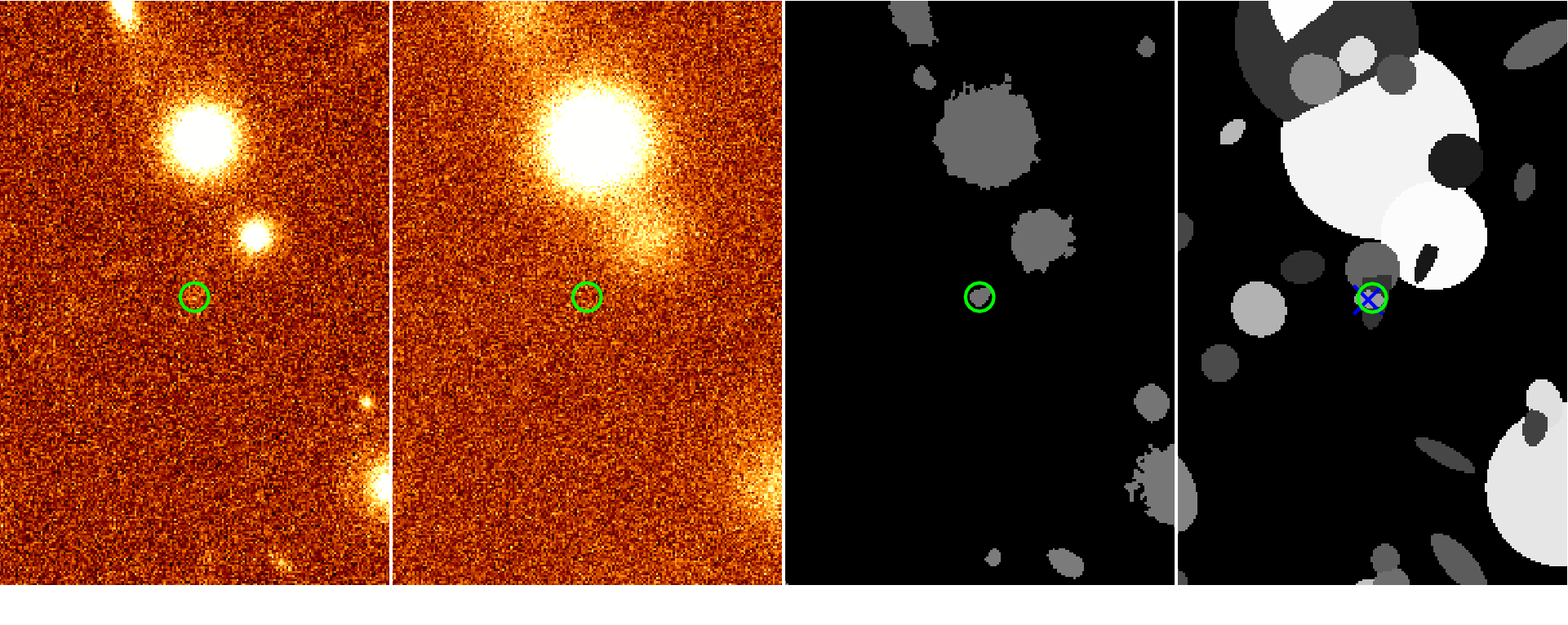}
\centering
\caption{Accuracy of the flux determination in a simulation 
containing extended objects, overlapping priors, and
\textsc{SExtractor} detection. Top: relative flux
difference $(f_{meas}-f_{true})/f_{true}$ versus logarithm of
the input flux $f_{true}$ for a simulated image populated with extended 
sources (FWHM=1.66"). 
Symbols and colours are as in Fig. \ref{accuracy1}.
The inner panel shows a magnification of the brighter end of the distribution.
The outlier marked with the open black circle, ID=720, is shown in the
bottom panel: left to right, HRI (FWHM=0.2"), LRI, 
\textsc{SExtractor}
segmentation map and ``true'' segmentation map. The green circles
show the object detected via \textsc{SExtractor}, while the blue cross
shows its ``true'' position. See text for more details.}\label{accuracy2}
\end{figure*}

\begin{figure*}[ht] 
\includegraphics[width=14cm]{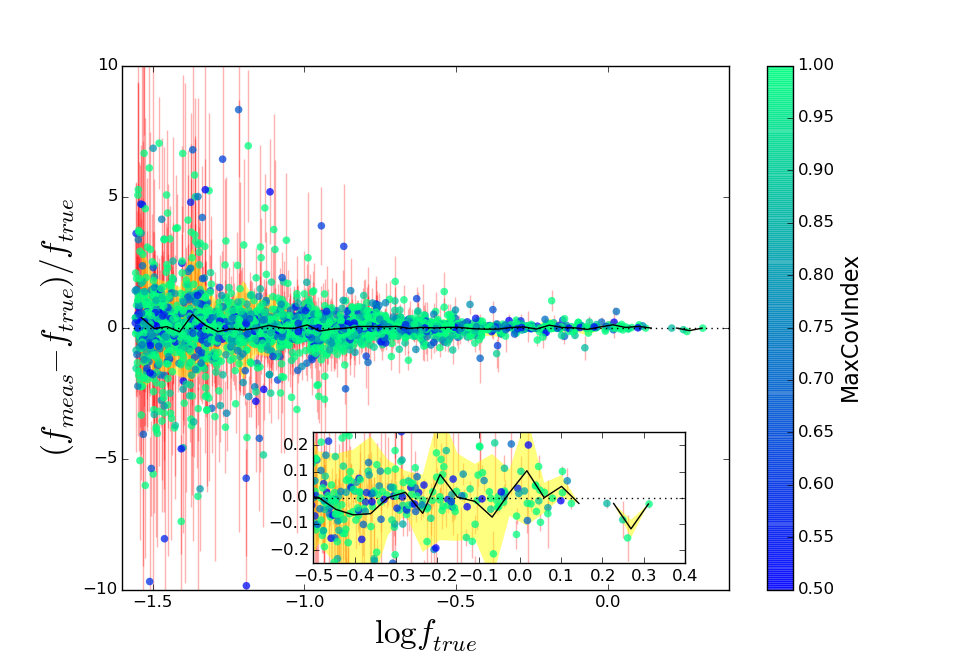}
\centering
\caption{Accuracy of the flux determination in a FIR-like
simulation (Herschel SPIRE 250 $\mu$m, FWHM=25'', 3.6'' pixel scale), 
using unresolved priors. The symbols have the same
meaning as in Fig. \ref{accuracy2}.  See text
for more details.}\label{GNsim}
\end{figure*}

\begin{figure*}
\includegraphics[width=9cm]{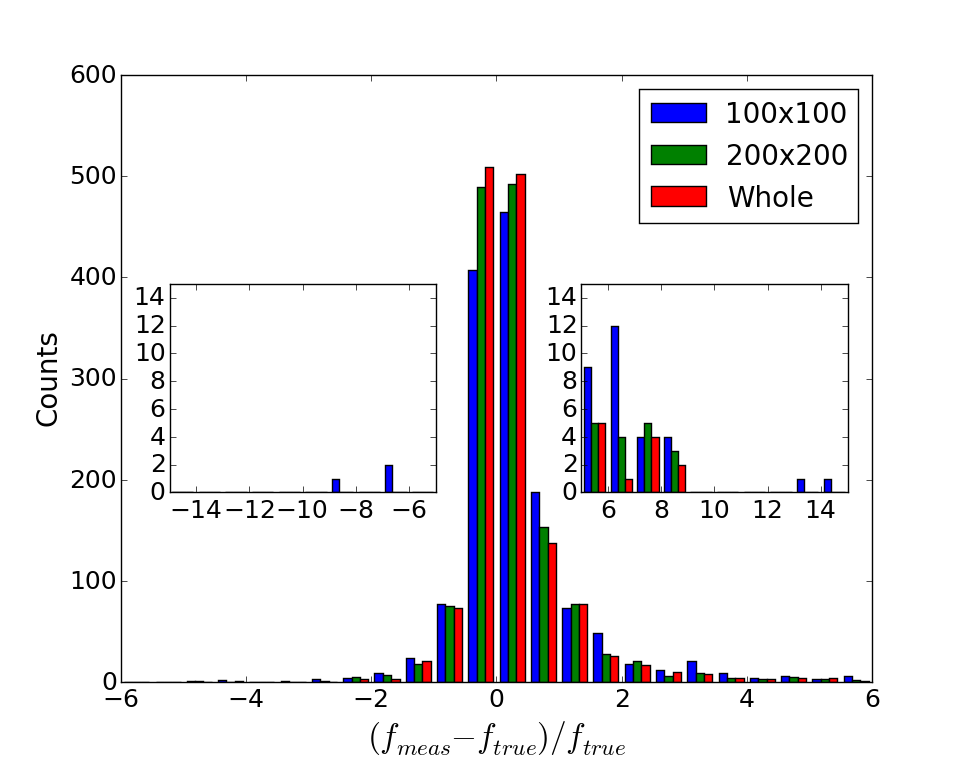}
\includegraphics[width=9cm]{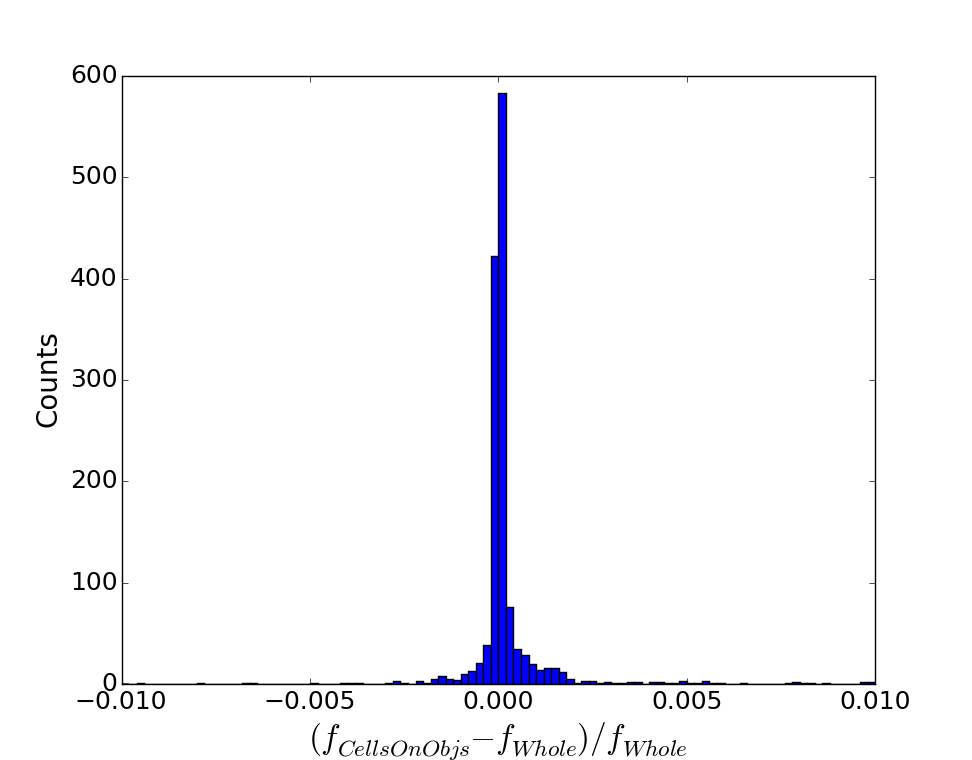}
\centering
\caption{Accuracy of the flux determination. Top panel: for the same simulation
described in Fig. \ref{accuracy2}, the histograms show the results for three
different fitting methods: regular grid 100$\times$100 pixels 
(standard \textsc{tfit} 
approach), regular grid 200$\times$200, single fit on the whole image. The small
boxes show the extended wings of the histograms, magnified for better viewing.
The accuracy increases by enlarging the cells and, reaches the best result with 
the single fit on the whole image.
Bottom panel: the histogram shows the relative measured 
flux difference between the single fit on the whole image and the
\textit{cells-on-objects} method. Differences above $1\%$ are very 
rare.}\label{accuracy3}
\end{figure*}

\begin{figure*}[ht] 
\includegraphics[width=16cm]{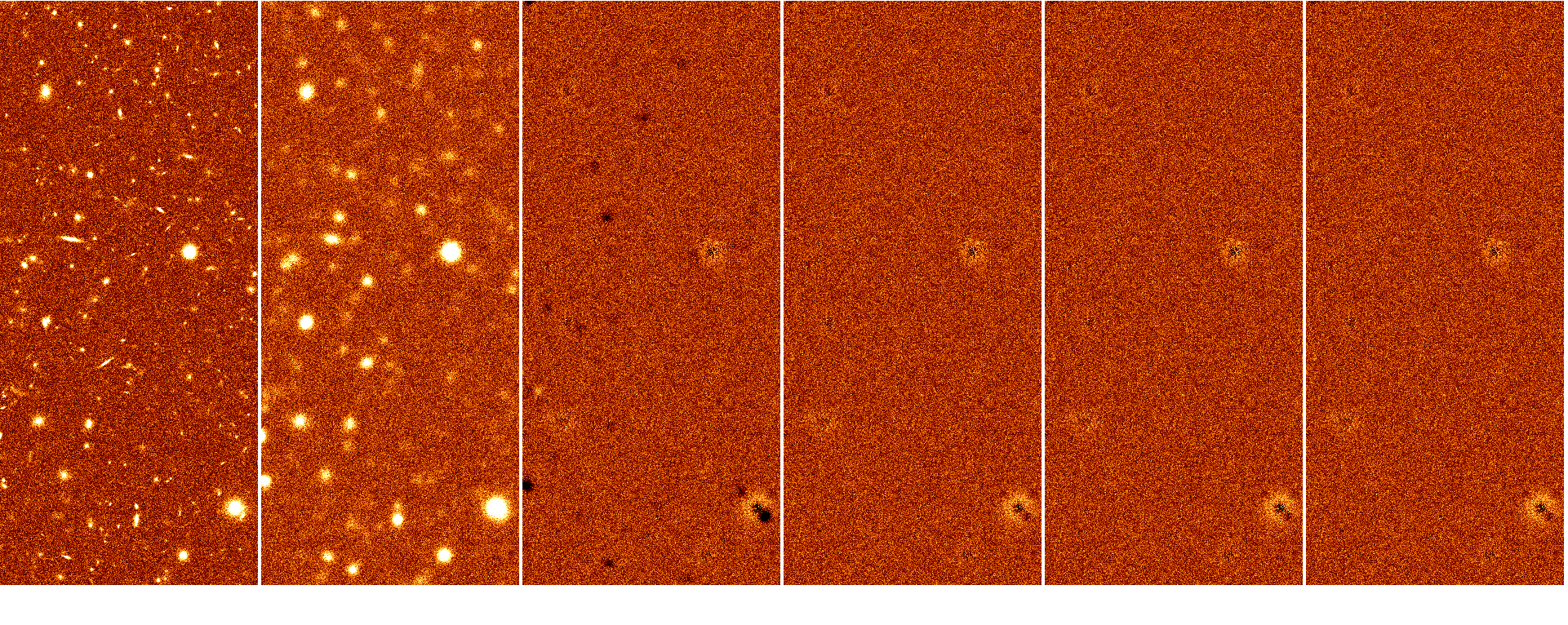}
\includegraphics[width=15.5cm]{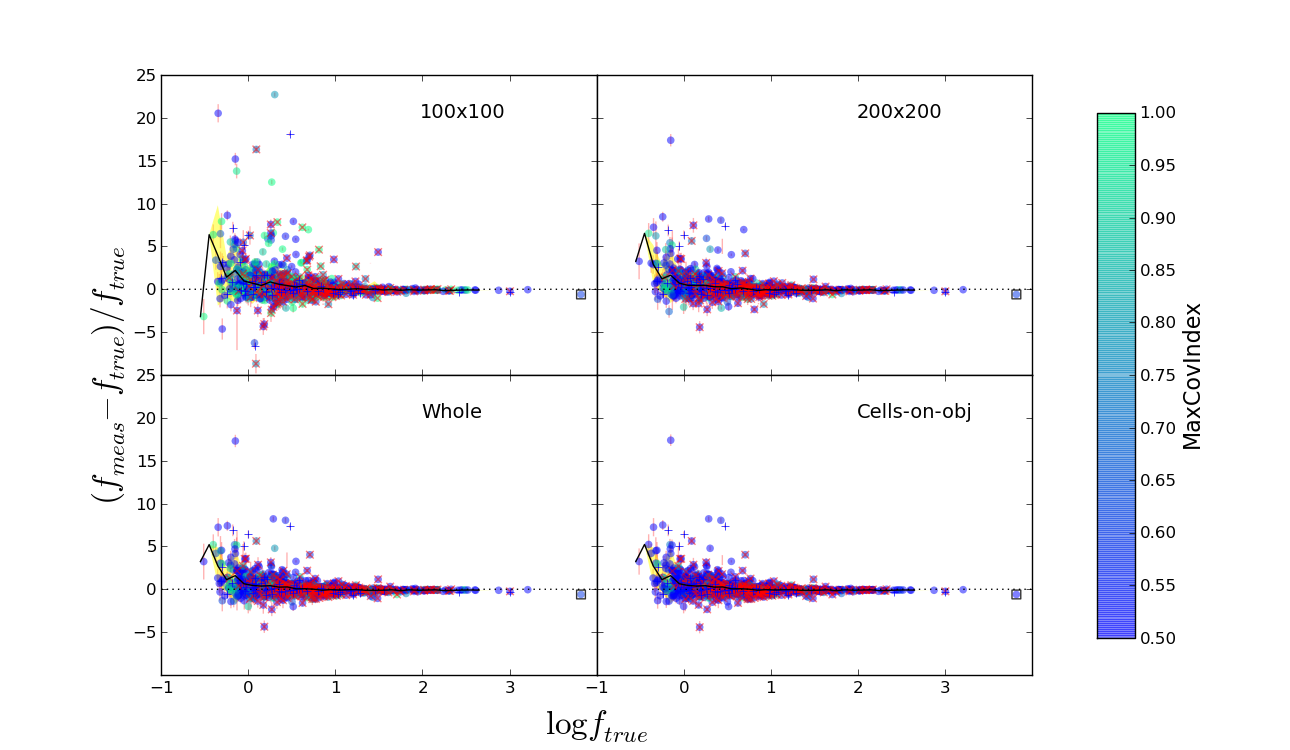}
\centering
\caption{Accuracy of the flux determination. For the same simulation
described in Fig. \ref{accuracy2}, the plots show the results for four
different fitting methods. Top panel, left to right: HRI 
(FWHM=0.2"), LRI (FWHM=1.66"), 
residuals using a regular grid of $100\times100$ 
pixels cells (standard \textsc{tfit} approach), 
a regular grid of $200\times200$ pixels cells, 
a single fit on the whole image, and the \textit{cells-on-objects} method.
The spurious fluctuations in the last two panels are due to segmentation 
inaccuracies, as in Fig. \ref{accuracy2}.
Bottom panels, left to right and top to bottom: relative measured flux
differences with respect to true fluxes, same order as above.
The values of the covariance index are different in each case
because of the varying sizes of the cells (and therefore of the relative
matrix).}\label{accuracy4}
\end{figure*}

In a subsequent more realistic test, we considered
extended objects (including morphologies of objects
from the \textsc{GenCat} catalogue, using 
FWHM$_{HRI}$=0.2" and 
FWHM$_{LRI}$=1.66" and imposing 
m$_{true,LRI}$=m$_{true,HRI}=m_{H160,GenCat}$ for simplicity) 
and allowed for overlapping priors. 
To be consistent with the standard procedure 
adopted for real images, for this 
case we proceeded by producing an \textsc{SExtractor} catalogue
and segmentation map, which were then spatially cross-correlated with the
``true'' input catalogue. The results for this test are shown in 
Fig. \ref{accuracy2}. Even in this much more complex situation, the
results are reassuring; there is an overall good agreement
between measured and input fluxes for bright ($\log S/N>1$) sources, 
with only a few flagged objects clearly showing large deviations from
the expected value, and a reasonably good average agreement
down to $\log S/N=0$. 
However, all bright fluxes are measured $\simeq 5\%$ 
fainter than the true values (see the inner box in the same figure); 
this is very likely the effect of the limited segmentation extension,
as already discussed in the previous section.
On the other hand, faint sources tend to have systematically overestimated 
fluxes, arguably because of contamination from undetected sources.
To confirm this, we focus our attention on a single case study (the 
source marked as ID 720) which shows a large discrepancy from 
its true flux, but has a relatively small covariance index. 
An analysis of the real segmentation map 
shows how the detected object is actually a superposition 
of two different sources that have been detected as a single one,
so that the measured flux is of course higher than expected.
One should also note that the uncertainties on the measured 
fluxes are smaller in this test, because there are fewer priors
(only the ones detected by \textsc{SExtractor} are now present), 
implying a lower rank of the covariance matrix and a lower number
of detected neighbours blending in the LRI. This causes 
a global underestimation of the errors.

To check the performance of \textsc{t-phot} at 
FIR wavelengths, we also run a test on a simulated Herschel 
SPIRE 250 $\mu$m image (FWHM=25'', 3.6'' pixel scale). 
The simulated image (shown alongside with the obtained residuals
in Fig. \ref{herschel}) mimics real images from the
GOODS-Herschel program, the deepest Herschel images ever obtained. This
image was produced with the technique presented in \citet{Leiton2015}; we
first derived (predicted) flux densities for all the 24 $\mu$m detections
($F_{24\mu m} > 20 \mu$Jy) in GOODS-North, which are dependent on their
redshift and flux densities at shorter wavelengths, and then we injected
these sources into the real noise maps from GOODS-Herschel imaging.
Additional positional uncertainties, typically 0.5$\arcsec$, were also applied
to mimic real images. As shown in \cite{Leiton2015}, these simulated
images have similar pixel value distributions to real images
\citep[see also][ for more details]{Wang2015}. For this test,
\textsc{t-phot} was run using the list of all the 
24 $\mu$m sources as unresolved priors. The results of the test
are plotted in Fig. \ref{GNsim}, and they show that even in this
case \textsc{t-phot} can recover the input fluxes
of the sources with great statistical accuracy (the mean of the
relative deviation from the expected measurements,
i.e. the black solid line in the plot, is consistent
with zero down to the faintest fluxes).  
The results are equivalent to those obtained 
on the same datasets with other public software 
specifically developed for FIR photometry,
such as \textsc{FastPhot} \citep{Bethermin2010}.


\subsubsection{Testing different fitting options: cell dimensions}

We then proceeded to check the performance of the different fitting 
techniques that can be used in \textsc{t-phot}. To this aim, we 
repeated the test on the 1.66'' LRI with extended priors
and \textsc{SExtractor} priors, described in Sect. 
\ref{realistic}, with different fitting methods:
using a regular grid of cells of $100\times100$ pixels, 
a regular grid of cells of $200\times200$ pixels, and the 
\textit{cells-on-objects} method, comparing the results 
with those from the fit of the whole image at once. 
The results of the tests are shown in Figs. \ref{accuracy3} 
and \ref{accuracy4}. The first figure compares the  
distributions of the relative errors in measured flux for 
the runs performed on the $100\times100$ pixels grid, on the 
$200\times200$ pixels grid, and on the whole image at once. 
Clearly, using any regular grid of cells worsens the results, 
as anticipated in Sect. \ref{news}. Enlarging the sizes of the cells
improves the situation, but does not completely solve the problem.
We note that the adoption of an arbitary grid of cells of any dimension 
in principle is prone to the introduction of potentially large errors, 
because (possibly bright) contaminating objects may
contribute to the brightness measured in the cell, 
without being included as contributing sources. A
mathematical sketch of this issue is explained in
the Appendix B (see also Sect. \ref{real}).
The second histogram compares the differences between the fit 
on the whole image and the fit with the \textit{cells-on-objects} 
method. Almost all the sources yield identical results with the 
two methods, within $(f_{meas}-f_{true})/f_{true}< 0.001$, which 
proves that the \textit{cells-on-objects} method can be 
considered a reliable alternative to the single-fit method.
Finally, Figure \ref{accuracy4} compares the 
HRI, the LRI, and the residual images obtained with the four 
methods and their distributions of relative errors, showing 
quantitatively the difference between the analyzed cases.

In summary, it is clear that an incautious choice of 
cell size may lead to unsatisfactory and catastrophic outcomes.
On the other hand, the advantages of using a single fit, 
and the equivalence of the results obtained with the 
single-fit and the \textit{cells-on-objects} techniques, are evident. 
As already anticipated, one should bear in mind that the 
\textit{cells-on-objects} method is only convenient if the overlapping
of sources is not dramatic, as in ground-based optical 
observations. For IRAC and FIR images, on the other hand,
the extreme blending of sources would cause the cells to
be extended over regions approaching the size of the whole 
image, so that a single fit would be more convenient, although
often still CPU-time consuming.

\begin{figure*}[ht] 
\includegraphics[width=16cm]{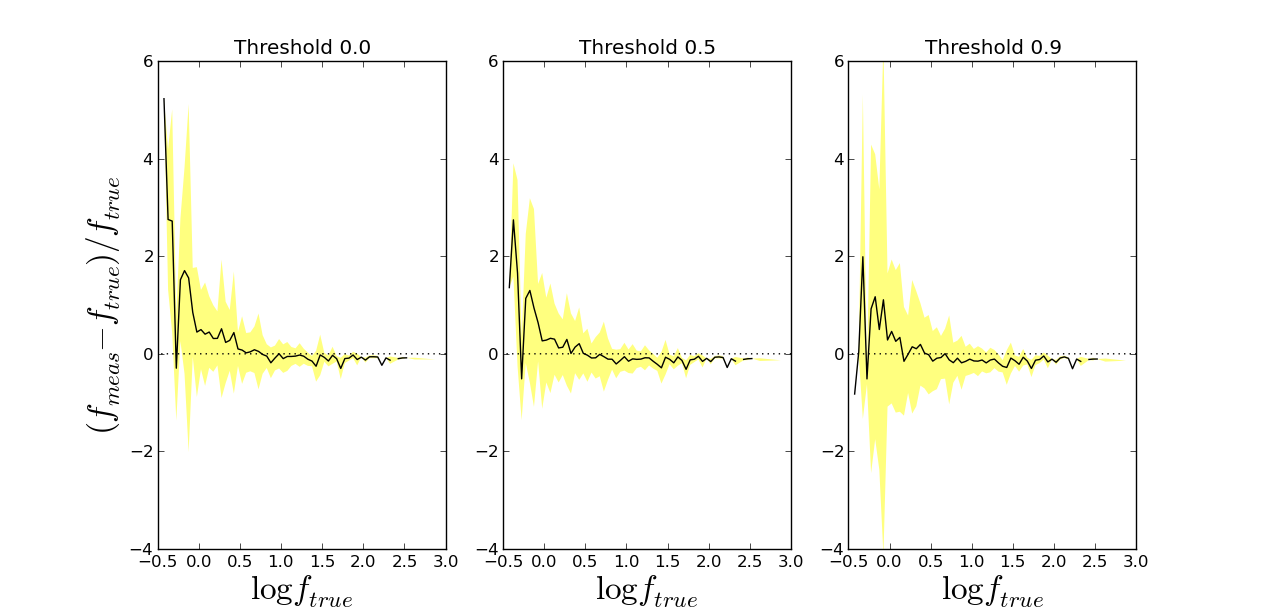}
\centering
\caption{Effects of threshold fitting
(Sect. \ref{threshold}). Mean relative error 
(black line) and standard deviation (yellow shaded area) for three
simulations with different threshold values (0.0, 0.5, 0.9). Only
pixels with normalized flux higher than the threshold values
are included in the fit. Larger threshold values result in
more accurate measurements for faint sources, at the expense
of a systematic underestimation of the flux for brighter 
ones.}\label{thresholdsaccuracy}
\end{figure*}

\begin{figure*}[ht]
\includegraphics[width=12cm]{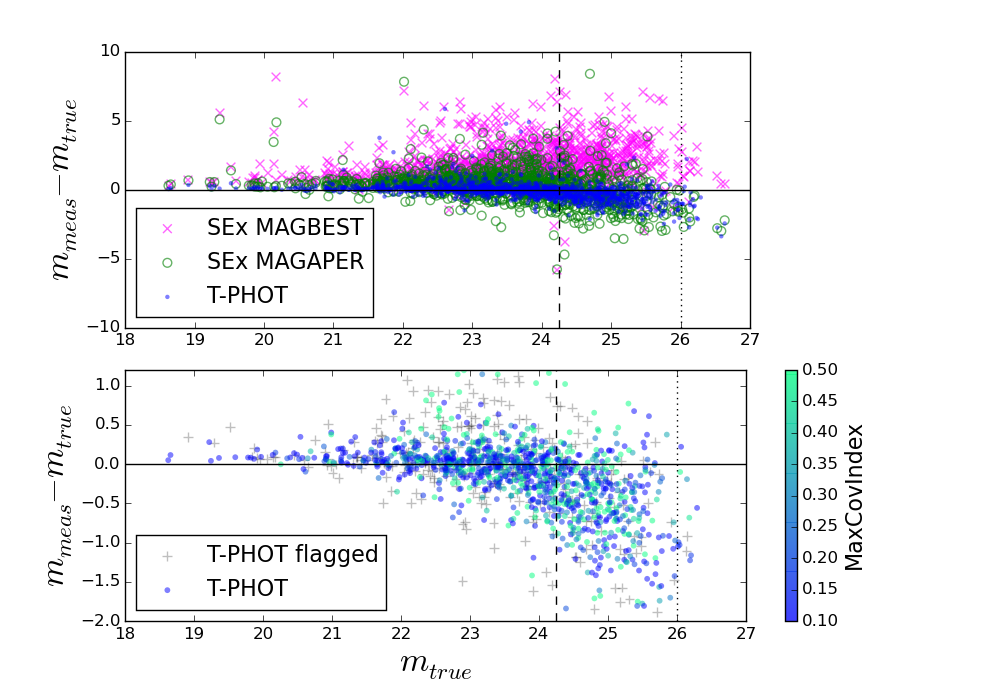}
\centering
\caption{Top: measured magnitude differences 
$(m_{meas}-m_{true})$ versus ``true'' input magnitudes 
$m_{true}$, for two simulated images 
populated with extended sources (HRI has FWHM=0.2" and HST 
$H$band-like fluxes, LRI has FWHM=1.66" and IRACch1-like 
fluxes), using three different methods:
\textsc{SExtractor} dual-mode aperture, \textsc{SExtractor}
dual-mode ``best'', and \textsc{t-phot}. Vertical lines show
the $5\sigma$ (dashed) and $1\sigma$ (dotted)
limits of the simulated LRI. Bottom:
magnification of the top panel, showing only
\textsc{t-phot} results, colour-coded as
a function of the covariance index.
See text for more details.}\label{colours1}
\end{figure*}

\subsubsection{Testing different fitting options: threshold 
fitting} \label{threshold}

As described in Sect. \ref{news}, \textsc{t-phot} includes
the option of imposing a lower threshold on the normalized
fluxes of templates so as to exclude low signal-to-noise
pixels from the fit. Figure \ref{thresholdsaccuracy} shows a 
comparison of the relative errors obtained with three 
different values of the \texttt{THRESHOLD} parameter: 
$t=0$, $t=0.5$, and $t=0.9$ (whic means that only
pixels with normalized flux $f_{norm}>t \times f_{peak}$ in the convolved 
template will be used in the fitting procedure). The differences
are quite small; however, a non-negligible global effect can be noticed:
all sources tend to slightly decrease their measurement of flux
when using a threshold limit. This brings faint sources (generally 
overestimated without using the threshold) closer to their true
value, at the same time making bright sources too faint. This
effect deserves careful investigation, which is beyond the scope of
this study, and is postponed to a future paper.

\subsubsection{Colours}

A final test was run introducing realistic colours,
i.e. assigning fluxes to the sources in the LRI consistent 
with a realistic SED (as output by \textsc{GenCat},
see Sect. \ref{realistic}),
instead of imposing them to be equal to the HRI fluxes. 
We took IRAC-ch1 as a reference filter for the LRI,
consistently with the chosen FWHM of 1.66''. Furthermore,
we allowed for variations in the bulge-to-disc
ratios of the sources to take possible effects
of colour gradients into account. 
We compared the results obtained with \textsc{t-phot} with
the ones obtained with two alternative methods
to determine the magnitudes of the sources in the LRI:
namely, \textsc{SExtractor} dual mode aperture
and \texttt{MAG\_BEST} photometry (with HRI as detection 
image).
The differences between measured and input magnitudes in the
LRI, m$_{meas}-$m$_{true}$, are plotted in Fig. 
\ref{colours1}. Clearly, \textsc{t-phot} ensures the best
results, with much less scatter in the measurements
than both of the other two methods, and very few outliers.

\subsection{Performance on real datasets} \label{real}

\begin{figure*}[ht] 
\includegraphics[width=9.5cm]{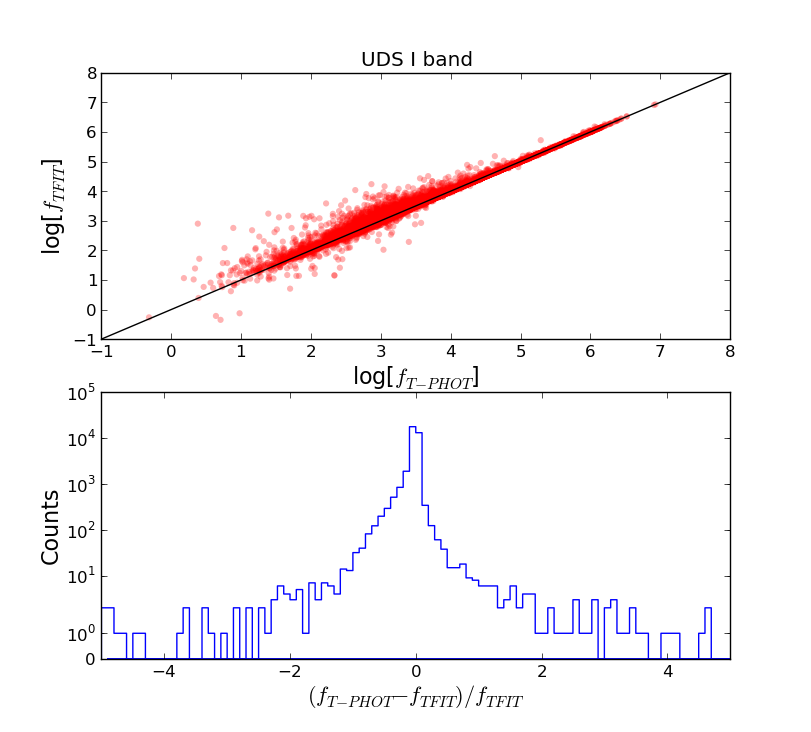}
\centering
\caption{UDS I band \textsc{tfit} versus \textsc{t-phot} comparison. Top
panel: compared measured fluxes. Bottom panel: histogram of relative
measured flux difference.}\label{uds1}
\end{figure*}

\begin{figure*}[ht] 
\includegraphics[width=12cm]{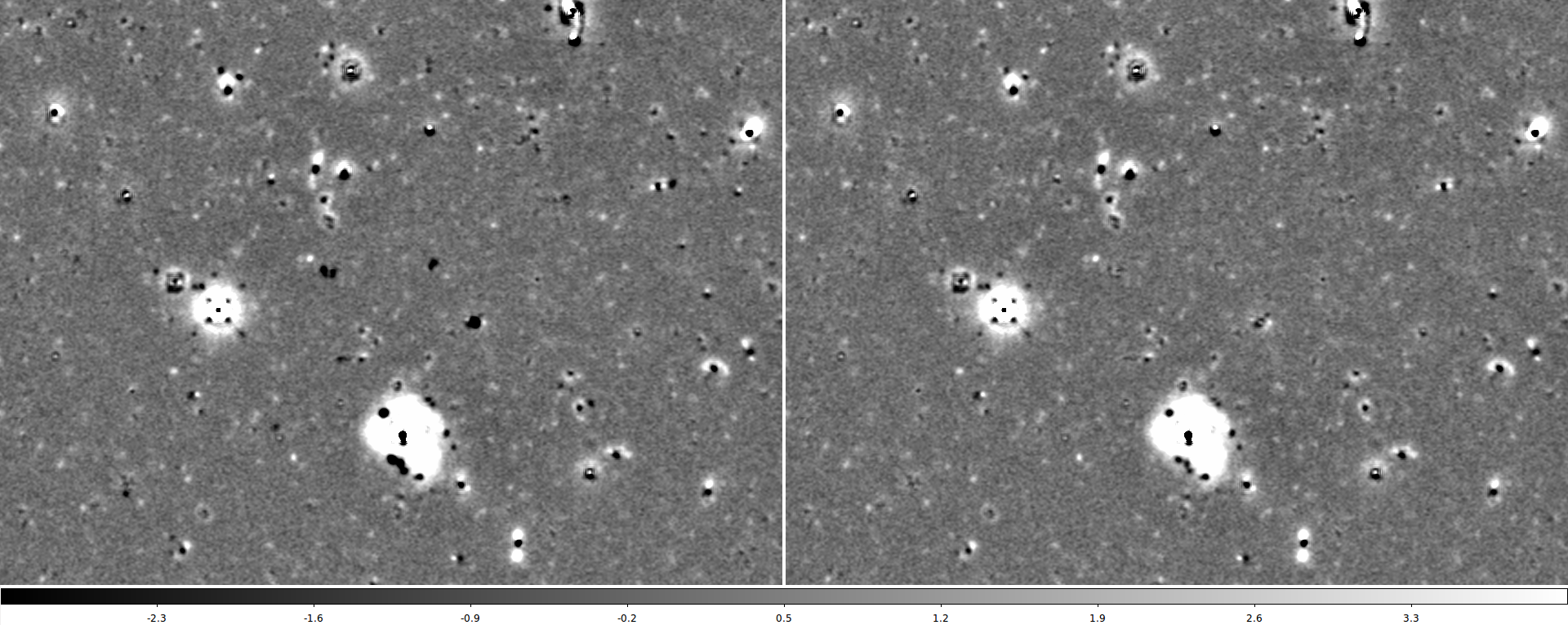}
\includegraphics[width=12cm]{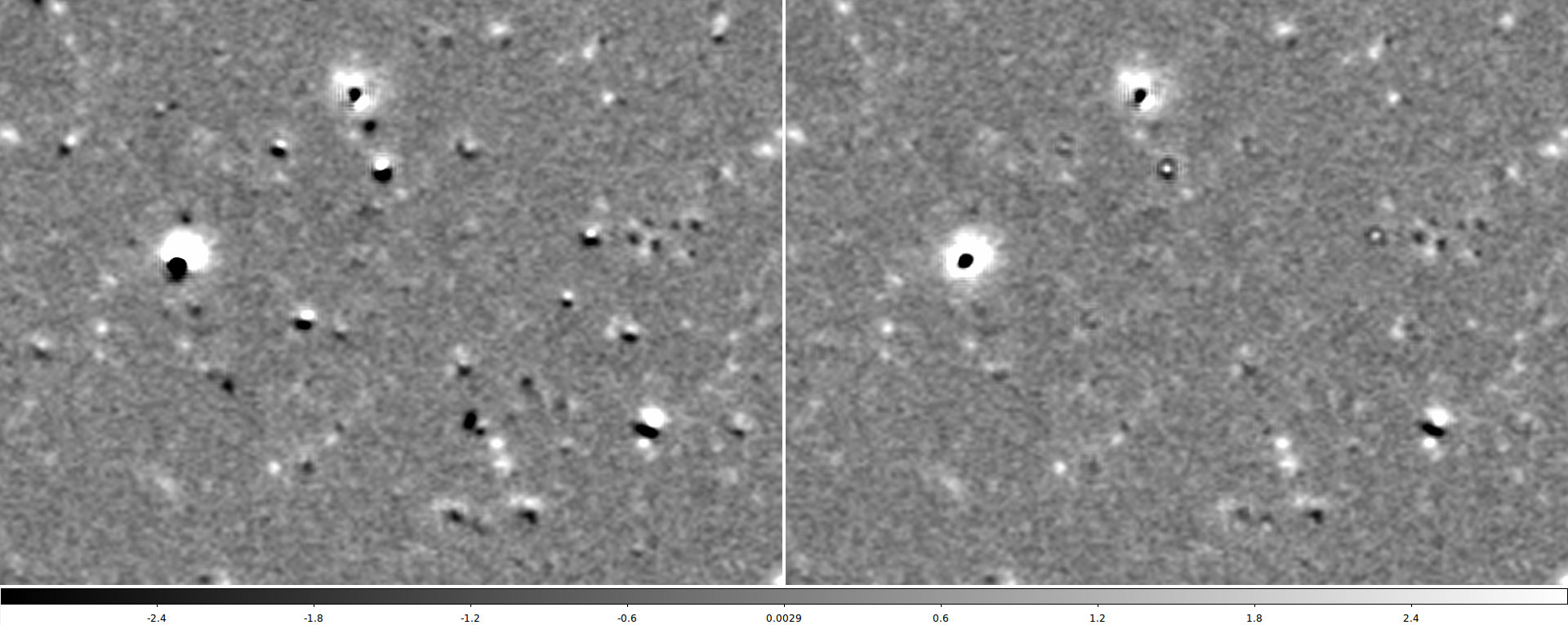}
\centering
\caption{UDS I band \textsc{tfit} versus \textsc{t-phot} comparison. The panels
on the left show two small patches of the official CANDELS residual image
obtained using \textsc{t-fit}. The residual images of the 
same regions are showed in the right panels, this time obtained 
using \textsc{t-phot} with \textit{cells-on-objects} method 
and improved local kernel registration.
We note the disappearence of many spurious black spots.}\label{uds2}
\end{figure*}

It is instructive to check how
\textsc{t-phot} performs on real datasets, in addition 
to simulations. To this aim, we run two different tests. In the first, 
we compared the results of the \textsc{tfit} CANDELS 
analysis on the UDS CANDELS $I$-band \citep{Galametz2013} 
to a \textsc{t-phot} run obtained  
using the \textit{cells-on-objects} method and different parameters
in the kernel registration stage.
Figure \ref{uds1} shows the histograms of the differences in the 
photometric measurements between \textsc{tfit} and \textsc{t-phot}.
Many sources end up with a substantially different flux, because
of the two cited factors (a better kernel registration and the 
different fitting procedure). We note that the majority of the sources
have fainter fluxes with respect to the previous measurements,
precisely because of the effect described in Sect. \ref{realistic}:
fitting using a grid of cells introduces systematic errors
assigning light from sources that are not listed in a given cell,
but overlap with it to the objects recognized as belonging to the cell.
To further check this point, Fig. \ref{uds2} shows some examples 
of the difference between the residuals obtained with \textsc{tfit} 
(official catalogue) and those obtained with this \textsc{t-phot} run 
using \textit{cells-on-objects} method, also introducing better 
registration parameters in the \texttt{dance} stage. Clearly, 
the results are substantially different, and many black spots
(sources with spurious overestimated fluxes) have disappeared. Also, 
the registrations appear to be  generally improved.

\begin{figure}[h!]
\includegraphics[width=8cm]{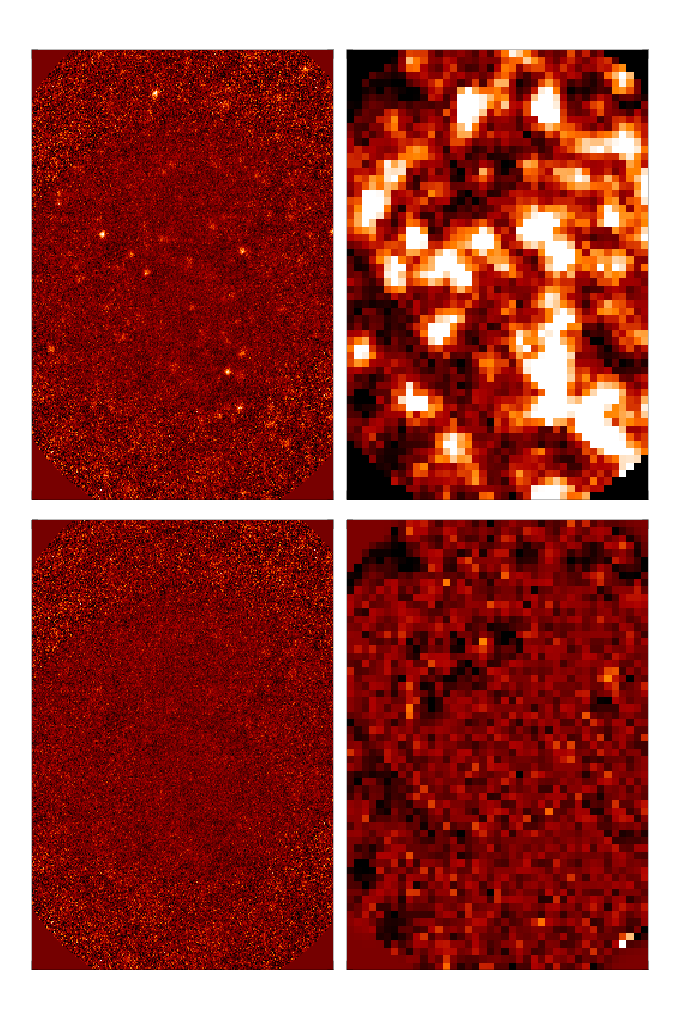}
\centering
\caption{Results from a test run using \textsc{t-phot} with unresolved
priors on FIR/sub-mm real dataset. Upper row, left to right:
SCUBA-2 450 $\mu$m (FWHM=7.5'') and Herschel 500 $\mu$m (FWHM=36'') 
images of the COSMOS-CANDELS fields. Lower row, left to right: 
residuals for the two fields, obtained with \textsc{t-phot} runs using a list
of 24+850 $\mu$m priors. See text for details.}\label{FIR_res}
\end{figure}

\begin{figure*}[ht]
\includegraphics[width=12cm]{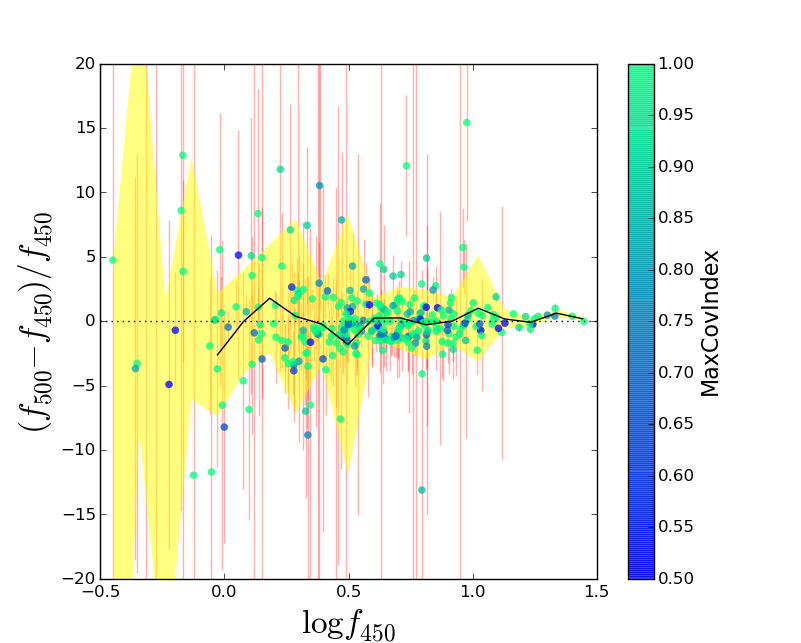}
\centering
\caption{Accuracy of the flux determination for the dataset
described in Fig. \ref{FIR_res}: measured Herschel 500 $\mu$m (FWHM=36'') 
fluxes $f_{500}$ are compared to the fluxes obtained for the SCUBA-2 450 
$\mu$m (FWHM=7.5'') $f_{450}$, considered as reference fluxes (using 
24+850 $\mu$m unresolved priors in both cases). The symbols have the 
same meaning as in Fig. \ref{accuracy2}; the error 
bars now include the measured error on the reference flux. 
See text for more details.}\label{FIR_diff}
\end{figure*}

The second test was run on FIR/sub-mm SCUBA-2 (450 $\mu$m,
FWHM=7.5'') and Herschel (500 $\mu$m, FWHM=36'') images of the COSMOS-CANDELS
field. In both cases, a list of 24+850 $\mu$m sources was used
as unresolved priors. Figure \ref{FIR_res} shows the original images
in the top row, and the residuals in the bottom row. The model has 
removed all significant sources from the 450 $\mu$m map and the majority
from the 500 $\mu$m map.
Figure \ref{FIR_diff} shows a comparison of the fluxes measured in 
the \textsc{t-phot} fits to the 450 $\mu$m and 500 $\mu$m maps 
at 24+850 $\mu$m prior positions, with the error bars combining the 
errors on both flux measurements. 
Agreement within the errors implies successful deconfusion of the 
Herschel image to reproduce the fluxes 
measured in the higher resolution SCUBA-2 image. 
This typology of analysis is very complex and we do not want to
address here the subtleties of the process; we refer the reader
to \citet[][ in preparation]{Wang2015} 
and \citet[][ in preparation]{Bourne2015} for detailed discussions
on the definition of a robust and reliable approach to measure FIR
and sub-mm fluxes. These simple tests, however, clearly show that
\textsc{t-phot} is successful at recovering the fluxes
of target sources even in cases of extreme confusion and blending,
within the accuracy limits of the method.

\section{Computational times} \label{times}

As anticipated, \textsc{t-phot} ensures a large saving of
computational time compared to similar codes like \textsc{tfit}
and \textsc{convphot} when used with identical input parameters.
For example, a complete, double-pass run on the whole CANDELS UDS field at once 
(I band; $\sim$35000 prior sources; LRI 30720$\times$12800 $\simeq$ 400 million 
pixels; standard \textsc{tfit} parameters and grid fitting) is completed without 
memory swaps in about 2 hours (i.e. 1 hour per pass) on a standard workstation
(\textsc{Intel i5}, 3.20 GHz, RAM 8 Gb).
A complete, double-pass run on the GOODS-S Hawk-I W1 field 
($\sim$17500 prior sources, LRI 10700$\times$10600 $\simeq$ 100 millions 
pixels, identical parameters) is completed in $\sim$20 minutes. 
For comparison, \textsc{tfit} may require many hours ($\sim$24) to complete 
a single pass on this Hawk-I field on the same machine. 
It must be said that \textsc{tfit} by default produces cutouts and templates 
for all the sources in the HRI image; selecting the ones belonging to the 
LRI field and inputting an ad hoc catalogue would have reduced the computing 
time by a factor of two (i.e. 11 hours for a single pass). 
It was not possible to process large images like the UDS field in a single 
run, because of RAM memory failure.
\textsc{convphot} timings and memory problems are similar to those of 
\textsc{tfit}, although they have different causes (being written in \texttt{C}, 
computation is generally faster, but it employs a slower convolution method 
and the solution of the linear system in performed as a single fit instead 
of grid fitting like in \textsc{tfit}, being much more time consuming).

Adopting the \textit{cells-on-objects} (Sect. \ref{news}) method increases the 
computational time with respect to the \textsc{tfit} standard cell approach,
but it is still far more convenient than the \textsc{convphot} standard 
single-fit approach, and gives nearly identical results.

Table \ref{tabtimes} summarizes the computational times for extended tests
on a set of simulated images having different detection depths 
(and therefore number of sources) and dimensions, with LRI FWHM=1.66''. 
The simulations were run on the same machine described above, using 
three different methods: whole image fitting, \textit{cells-on-objects},
and $100\times100$ pixels cells fitting.

\begin{table*}[t!]
\begin{center}
  \begin{tabular}{ | l | l | l | l | }
    \hline
    \backslashbox{Size[pix]}{mag$_{lim,det}$} & 27 & 28 & 29 \\ 
    \hline
    \multicolumn{4}{|c|}{\textbf{Number of sources}}\\
    \hline 
    $2500\times2500$ & 523 & 1070 & 1398 \\ \hline
    $5000\times5000$ & 2104 & 4237 & 5561 \\ \hline
    $10000\times10000$ & 8390 & 16807 & 22394 \\ \hline
    $20000\times20000$ & 33853 & 65536 & 65536 \\ \hline
    \hline
    \multicolumn{4}{|c|}{\textbf{Whole image fitting}} \\
    \hline 
    $2500\times2500$ & 38'' (2'') & 54'' (10'') & 1'9'' (20'') \\ \hline
    $5000\times5000$ & 3'26'' (1'1'') & 11'9'' (7'41'') & 20'28'' (16'1'') \\ \hline
    $10000\times10000$ & 1h28'22'' (1h15'46'') & 8h26'1'' (7h58'10'') & 21h16'24'' (20h27'53'') \\ \hline
    $20000\times20000$ & - & - & - \\ \hline
    \hline
    \multicolumn{4}{|c|}{\textbf{\textit{Cells-on-objects} fitting}} \\
    \hline
    $2500\times2500$ & 46'' (4'') & 1'11'' (16'') & 1'30'' (33'') \\ \hline
    $5000\times5000$ & 3'1'' (18'') & 4'27'' (1'8'') & 6'3'' (2'20'') \\ \hline
    $10000\times10000$ & 12'27'' (1'12'') & 17'52'' (4'31'') & 25'11'' (9'52'') \\ \hline
    $20000\times20000$ & 51'12'' (6'1'') & 1h34'40'' (35'8'') & 1h43'10'' (41'2'') \\ \hline
    \hline
    \multicolumn{4}{|c|}{\textbf{100$\times$100 pixels cells fitting}} \\
    \hline
    $2500\times2500$ & 52'' (3'') & 1'6'' (7'') & 1'14'' (9'') \\ \hline
    $5000\times5000$ & 3'16'' (14'') & 4'22'' (29'') & 4'54'' (41'') \\ \hline
    $10000\times10000$ & 13'4'' (56'') & 17'12'' (1'54'') & 19'47'' (2'53'') \\ \hline
    $20000\times20000$ & 55'24'' (6'19'') & 1h18'38'' (15'53'') & 1h17'17'' (17'58'') \\ \hline
  \end{tabular}
\end{center}
\caption{Test of computational times for \textsc{t-phot} runs on images 
of given dimensions and limiting magnitude in detection. 
Each entry of the table gives the total duration of run, 
the duration the fitting stage alone (in parentheses),
and the number of fitted sources. The \texttt{dance} stage takes
most of the CPU time after the fitting routine.} \label{tabtimes}
\end{table*}

\section{Summary and conclusions} \label{conclusions}

We have presented \textsc{t-phot}, a new software package developed
within the \textsc{astrodeep} project. \textsc{t-phot}
is a robust and versatile tool, aimed at the photometric analysis
of deep extragalactic fields at different wavelengths and spatial
resolution, deconfusing blended sources in low-resolution images.

\textsc{t-phot} uses priors obtained from a high-resolution detection
image to obtain normalized templates at the lower resolution
of a measurement image, and minimizes a $\chi^2$ problem to
retrieve the multiplicative factor relative to each source,
which is the searched quantity, i.e. the flux in the LRI. 
The priors can be either real cutouts from the HRI,
or a list of positions to be fitted as PSF-shaped sources, or
analytical 2-D models, or a mix of the three types. Different options
for the fitting stage are given, including a \textit{cells-on-objects} 
method, which is computationally efficient while yielding accurate 
results for relatively small FWHMs. 
\textsc{t-phot} ensures a large saving of computational time 
as well as increased robustness with respect to similar 
public codes like its direct predecessors \textsc{tfit}
and \textsc{convphot}. With an appropriate choice of the parameter
settings, greater accuracy is also achieved.

As a final remark, it should be pointed out that the
analysis presented in this work deals with 
idealized situations, namely simulations or comparisons
with the performances of other codes on real datasets. There are
a number of subtle issues regarding complex aspects of the 
PSF-matching techinque, which become of crucial importance
when working on real data. A simple foretaste of such
complexity can be obtained by considering the problem 
described in Sect. \ref{validation}, i.e. the correct 
amplitude to be assigned to the segmented area of a 
source. Work on this is ongoing, and the full discussion will be 
presented in a subsequent companion paper.

As we have shown, \textsc{t-phot} is an efficient tool for
the photometric measurements of images on a very broad range of 
wavelengths, from UV to sub-mm, and is currently being routinely 
used by the \textsc{Astrodeep} community to
analyse data from different surveys (e.g. CANDELS, Frontier Fields, 
AEGIS).
Its main advantages with respect to similar codes like
\textsc{tfit} or \textsc{convphot} can be summarized as follows:
\begin{itemize}
\item when used with the same parameter settings of \textsc{tfit},
\textsc{t-phot} is many times faster (up to hundreds of times),
and the same can be said with respect to other similar codes
(e.g., \textsc{convphot});
\item \textsc{t-phot} is more robust, more user-friendly, and can
handle larger datasets thanks to an appropriate usage of the RAM;
\item \textsc{t-phot} can be used with three different types of
priors (real high-resolution cutouts, analytical models and/or
unresolved point sources) making it a versatile tool for the 
analysis of different datasets over a wide range of wavelengths
from UV to sub-mm;
\item \textsc{t-phot} offers many options for performing the
fit in different ways, and with an appropriate choice of 
parameter settings it can give more accurate results.
\end{itemize}

Future applications might include the processing of
EUCLID and CCAT data. New releases of the software package,
including further improvements and additional options,
are planned for the near future.

\begin{acknowledgements}
The authors acknowledge the contribution of the FP7 SPACE project “ASTRODEEP” (Ref.No: 312725), supported by the European Commission.
JSD acknowledges the support of the European Research Council via the award of an Advanced Grant.
FB acknowledges support by FCT via the postdoctoral fellowship SFRH/BPD/103958/2014 and also the funding from the programme UID/FIS/04434/2013.
RJM acknowledges the support of the European Research Council
via the award of a Consolidator Grant (PI McLure).
The authors would like to thank Kuang-Han Huang, Mimi Song, Alice Mortlock and Michal Michalowski, for constructive help and suggestions, and the anonymous referee for useful advice.
\end{acknowledgements}

\bibliographystyle{aa}
\bibliography{mnemonic,biblio}

\begin{appendix}

\section{The parameter file}

Below is a template of the standard first-pass parameter file 
to be given as input to \textsc{t-phot} (similar templates 
for both the first and the second pass are included
in the dowloadable tarball). It is very similar to the
original \textsc{tfit} parameter file, and part of the
description is directly inherited from it.

\begin{figure}[h!]
\centering
\begin{verbatim}
# T-PHOT PARAMETER FILE

# PIPELINE 

# 1st pass
order standard
#priors, convolve, fit, diags, dance, plotdance 

# PRIORS STAGE

# Choose priors types in use:
usereal	 	True  
usemodels	True  
useunresolved	True  

# Real 2-d profiles
hiresfile       HRI.fits 
hirescat        HRI.cat	 
hiresseg        HRI.seg.fits 
normalize       true	     
subbckg		True         
savecut		true	     
cutoutdir       cutouts	     
cutoutcat       cutouts/_cutouts.cat 

# Analytical 2-d models
modelscat	models/models.cat  
modelsdir	models		   

culling		false 

# Unresolved point-like sources
poscat		pos.cat 
psffile 	psf.fits        

# CONVOLUTION STAGE

loresfile       LRI.fits  
loreserr        LRI.rms.fits 
errtype         rms
rmsconstant     1 
relscale        1 

FFTconv		true 
multikernels	false	
kernelfile      kernel.fits
kernellookup    ch1_dancecard.txt 

templatedir     templates      
templatecat	templates/_templates.cat 

# FITTING STAGE

# Filenames:
fitpars	         tpipe_tphot.param
tphotcat         lores_tphot.cat_pass1
tphotcell        lores_tphot.cell_pass1
tphotcovar       lores_tphot.covar_pass1
\end{verbatim}
\end{figure}

\begin{figure}[h!]
\centering
\begin{verbatim}
# Control parameters:
fitting		coo 
cellmask        true 
maskfloor       1e-9 
writecovar      true 

threshold	0.0 
linsyssolver	lu 
clip		false 

# DIAGNOSTICS STAGES 

modelfile       lores_collage_pass1.fits 

# Dance:	
dzonesize       100 
maxshift        1.0 
ddiagfile       ddiags.txt 
dlogfile        dlog.txt 
dancefft	false 

\end{verbatim}
\end{figure}

\subsection{Pipeline}
Standard optical/NIR double-pass runs can be achieved 
by setting \texttt{order standard} and \texttt{order standard2}.

A standard first-pass run includes the stages
\texttt{priors, convolve, fit, diags, dance, plotdance}.
The stage \texttt{priors} allows for an automatic re-construction
of the pipeline depending on the input data given
in the following sections (see the documentation
included in the tarball). A standard second-pass run
includes the stages \texttt{convolve, fit, diags, archive}.
The \texttt{archive} stage creates a directory after the name of the LRI, 
with some specifications, and archives the products of both runs.

Double-pass runs for FIR/sub-mm can be achieved by setting
\texttt{order positions, fit, diags, dance, plotdance}
and \texttt{order positions, fit, diags, archive}.

\subsection{Priors}
Each prior must have a unique identification number
(ID) to avoid errors. The user must be careful to give
the correct information in this paramfile.
Select the priors to be used by switching
on/off the relative keywords: \texttt{usereal}, 
\texttt{usemodels}, \texttt{useunresolved}.

\begin{itemize}
\item \texttt{hiresfile}: the high-resolution, detection image.
If a catalogue and a segmentation map are given in the two
subsequent entries (\texttt{hirescat} and \texttt{hiresseg}), 
cutouts will be created out of this image. 
This step is necessary if a catalogue of real or model priors are
to be used. The catalogue \texttt{hirescat} must be in a standard
format: \texttt{id x y xmin ymin xmax ymax background SEx\_flux}
(\texttt{x} and \texttt{y} are the coordinates of the source in
HRI pixel reference frame; \texttt{xmin, ymin, xmax, ymax} are
the limits of the segmentation relative to the source in
HRI pixel reference frame; \texttt{background} is the value of
the local background; and \texttt{SEx\_flux} is a reference
isophotal flux). 
\item \texttt{poscat}: a catalogue of positions for unresolved,
point-like sources. No HRI image/segmentation is needed,
while the PSF to be used to create the models is mandatory
(\texttt{psffile}).
The catalogue must be in the standard format \texttt{id x y}.
\item \texttt{modelscat}: a catalogue (with format
\texttt{id x y xmin ymin xmax ymax background SEx\_flux},
as for a standard HRI priors catalogue) of model priors. 
\texttt{modelsdir} is the directory in which the stamps
of the models are stored. 
Models with two or more components can be processed, but
each component must be treated as a separated object,
with a different ID, and a catalogue for each component must
be given. Catalogues for each component must have the same name, 
but ending with "\_1", "\_2", etc.; put the "\_1" catalogue
in the paramfile. \emph{It is important to note that two components of 
the same object should not have exactly identical positions,
to avoid numerical divergencies}.
\item \texttt{culling}: if \texttt{True}, objects in the catalogue 
(real priors and/or models) 
but not falling into the LRI frame will not be processed; if
it is false, all objects in the catalogue will be processed
(useful for storing cutouts for future reuse on different
datasets) and the selection of objects will be done before
the convolution stage.
\item \texttt{subbckg}: if \texttt{True}, subtract the value 
given in the input catalogue from each cutout stamp.
\item \texttt{cutoutdir}: the directory containing the cutouts.
\item \texttt{cutoutcat}: the catalogue of the cutouts, containing
the flux measured within the cutout area (which may be different
from the \texttt{SEx\_flux} given in the input catalogue, e.g. if
the segmentation has been dilated).
We note that these are output parameters if you start from the 
\texttt{priors/cutout} stage; they are input parameters for 
the convolve stage.
\item \texttt{normalize}: determines whether the cutouts will be
normalized or not; it is normally set to true, so that the
final output catalogue will contain fluxes rather than colours.
\end{itemize}

\subsection{Convolution}
\begin{itemize}
\item \texttt{loresfile, loreserr}: the LRI and RMS images.
\textsc{t-phot} is designed to work with an RMS map as the error map,
but it will also accept a weight map, or a single constant
value of the RMS from which an RMS map is generated. The
errtype specifies which kind of error image is provided.
For best results, use a source-weighted RMS map, to prevent
the bright objects from dominating the fit.
\item \texttt{relscale}: the relative pixel scale between the 
two images. For example if the HRI has a pixel scale 
of 0.1 arcsec/pixel and the LRI has a pixel scale of 0.5
arcsec/pixel, the value of relscale should be 5. If the 
LRI has been manipulated to match the HRI pixel scale 
and WCS data (e.g. using codes like \textsc{Swarp} by 
E. Bertin), put \texttt{relscale 1}.
\item \texttt{kernelfile}: the convolution kernel file.
The kernel must be a FITS image on the same pixel scale 
as the high-resolutuion image. It should contain a centred, normalized image.
\item \texttt{FFTconv} is \texttt{True} if the convolution of cutouts with
the smoothing kernel is to be done in Fourier space (via \textsc{FFTW3}).
\item \texttt{kerntxt} may be explicitely put \texttt{True} if one wishes to 
use a text file containing the kernel instead of a \texttt{.fits} 
one. \textsc{t-phot} supports the use of multiple kernels to accommodate
a spatially varying PSF. To use this option, set the
\texttt{multikernels} value to true, and provide a \texttt{kernellookup} 
file (it is automatically produced during the \texttt{dance} stage
in the first pass, but it can also be fed externally)
that divides the LRI into rectangular zones,
specified as pixel ranges, and provides a local convolution 
kernel filename for each zone. Any object that falls in a zone
not included in the lookup file will use the transfer kernel
specified as kernelfile.
\item \texttt{templatedir}: the directory containing the templates created
in the convolve stage, listed in the catalogue \texttt{templatecat}.
We note that these are output parameters for the convolve stage,
and an input parameter for all subsequent stages.
\end{itemize}

\subsection{Fitting stage}
\begin{itemize}
\item \texttt{fitpars, tphotcat, tphotcell, tphotcovar}: 
these are all output parameters. The \texttt{tfitpars} file
specifies the name of the special parameter file for the fitting
stage that will be generated from the parameters in this file.
The others are filenames for the output catalogue, cell, and
covariance files, respectively.
\item \texttt{fitting}: this keyword tells \textsc{t-phot} which method to use
to perform the fitting (see also Appendix B):
\begin{itemize}
\item \texttt{coo} or \texttt{0} for \textit{cells-on-objects}; 
\item \texttt{single} or \texttt{-1} for single fit; 
\item \texttt{single!} or \texttt{-10} for optimized single fit (the LRI
is divided in square cells containing roughly 10000 sources each); 
\item \texttt{cell\_xdim, cell\_ydim, cell\_overlap} for an arbitrary grid of cells.
\end{itemize}
\item \texttt{cellmask}: if \texttt{true}, it uses a mask to exclude 
pixels from the fit that do not contain a value of at least 
\texttt{maskfloor} in at least one template. 
\item \texttt{writecovar}: if \texttt{true}, it writes the covariance 
information out to the \texttt{tphotcovar} file.
\item \texttt{threshold}: forces the use of a threshold on
the flux, so that only the central parts of the objects 
are used in the fitting process.
\item \texttt{linsyssolver}: the chosen solution method, 
i.e. LU, Cholesky, or Iterative Biconjugate Gradient (IBG). 
LU is default.
\item \texttt{clip}: tells whether to loop on the sources 
excluding negative solutions.
\end{itemize}

\subsection{Diagnostic stages}
\begin{itemize}
\item \texttt{modelfile}: the \texttt{.fits} file that will contain 
the collage made by multiplying each template by its best flux 
and dropping it into the
right place. An additional diagnostic file will be created: it will
contain the difference image (LRI - \texttt{modelfile}). Its filename
will be created by prepending \texttt{resid\_} to the modelfile.
\item \texttt{dzonesize} specifies the size of the rectangular zones over 
which the pixels' cross-correlation between LRI and \texttt{modelfile}
will be calculated during the \texttt{dance} stage. 
It should be comparable to the size over which 
misregistration should be roughly constant, but it must be
large enough to contain enough objects to provide a good 
signal to the cross-correlation.
\item \texttt{maxshift} specifies the maximum size of 
the \texttt{x,y} shift, in LRI pixel frame, that is considered 
to be valid. Any shift larger than this is considered spurious 
and dropped from the final results, and replaced by an interpolated 
value from the surrounding zones. Ideally, \texttt{maxshift} 
$\simeq 1 \mbox{pixel} \times \mbox{FWHM}_{LRI}$/$\mbox{FWHM}_{HRI}$.
\item \texttt{ddiagfile} is an output parameter for the \texttt{dance} 
stage, and an input parameter for the \texttt{plotdance} stage.
\item \texttt{dlogfile} is an output parameter; it simply contains 
the output from the cross-correlation process.
\item \texttt{danceFFT}: if \texttt{True} cross-correlation is to be 
performed using FFT techniques rather than in real pixel space.
\end{itemize}

\section{The cells-on-objects algorithm}

Experiments on simulated images (see Sect. \ref{validation})
clearly show that fitting small regions (cells) of the LRI, 
as done by default in \textsc{tfit}, may potentially lead to 
large errors. This is particularly true if the dimensions
of the cells are chosen to be smaller than an ideal size, which changes from 
case to case, but which should always be greater than
$\sim$10 times the FWHM. However, it can be mathematically shown that the 
``arbitrary cells'' method intrinsically causes the introduction 
of errors in the fit, as soon as a source is excluded from the cell 
(e.g. because its centre is outside the cell), but contributes 
with some flux in some of its pixels.

Consider a cell containing $N$ sources. 
For simplicity, assume that each source $i$ only overlaps with the 
two neighbours $i-1$ and $i+1$. Furthermore, assume that a $(N+1)$-th 
source is contaminating the $N$-th source, but is excluded from 
the cell for some reason, for example (as in \textsc{tfit}) because
the centroid of the source lies outside the cell.

The linear system for this cell $AF=B$ will consist of a matrix $A$
with only the elements on the diagonal and those with a $\pm 1$ offset 
as non-zero elements (a symmetric \textit{band} matrix), 
and the vector $B$ will contain the products of templates
of each source with the real flux in the LRI (as a summation on all pixels),
as described in Sect. \ref{linsys}.
Given the above assumptions, this means that the $N$-th term of $B$ will
be higher than it should be (because it is contaminated by the external source).

Using the Cramer rule for the solution of squared linear systems, the
flux for the object $i$ is given by
\begin{equation}
f_i = \frac{\det A_i}{\det A}
\end{equation}

with $A_i$ a square matrix in which the $i$-th columns is substituted with
the vector $B$. If for example $N=3$, for $i=1$ this gives
\begin{equation}
f_1 = [B_1 (A_{22} A_{33} - A_{23}^2) - A_{12}(B_2 A_{23} - B_3 A_{22})] / \det A
\end{equation}
and since $B_3$ is larger than it should be, $f_1$ will be overestimated 
(slightly, if $A_{12}$ is not large, i.e. if sources 1 and 2 do not strongly 
overlap). On the other hand, for $i=3$ we have
\begin{equation}
f_3 = [A_{11} (A_{22} B_3 - A_{23} B_2) - A_{12}^2 B_3 + B_1 (A_{12} A_{23})] / \det A
\end{equation}
and in this case again $A_{12}^2$ might be small, but the first term given
by $A_{11}A_{22}$ will certainly be large, resulting in a catastrophic 
overestimation of $f_3$. The value of $f_2$ will of course be underestimated, as it
would be easy to show.

From this simple test case it is clear that arbitrarily dividing the LRI
into regions will always introduce errors (potentially non-negligible) in
the fitting procedure, unless some method for removing dangerous contaminating
sources is devised.

The \textit{cells-on-objects} algorithm aims at ensuring the accuracy of
the flux estimate while at the same time drastically decreasing computational
times and memory requirements. As explained in Sect. \ref{linsys}, when this method
is adopted a cell is centred around each detected source, and enlarged to include
all its ``potential'' contaminant neighbours, and the contaminant of the contaminants,
and so on. To avoid an infinite loop, the process of inclusion is interrupted when one of 
the following criteria is satisfied:
\begin{itemize}
\item the flux of the new neighbour is lower than a given fraction $f_{flux}$ of the 
flux of the central object (the considered fluxes are: if real priors are used, 
the ones given in the HRI catalogue; if unresolved priors are used, the ones read 
in the pixels of the LRI containing the coordinates of the sources; if analytical 
models are used, the ones of the models as reported in the HRI models catalogue), 
or 
\item the template of the neighbour overlaps with its direct previous contaminant
for a fractional area lower than $f_{area}$.
\end{itemize}
Experiments on simulations have shown that good results are obtained with $f_{flux}=0.9$
and $f_{area}=0.25$, and these values are used as constants in the source code.

We note that if a cell is enlarged to more than 75\% of the dimensions of the total LRI,
\textsc{t-phot} automatically switches to the single fit on the whole image.

\section{Suggested best options}

Of course, different problems require different approaches in 
order to obtain their best possible solution, and users are encouraged 
to try different options and settings. However, some indicative guidelines 
for optimizing a run with \textsc{t-phot} can be summarized as follows.
\begin{itemize}
\item Be sure that all the required input files exist and
have correct format, and that paths are correctly given
in the parameter file.
\item Whenever possible, fit the whole image at once (i.e. put
\texttt{fitting single} in the parameter file). The more sources there are,
and the more severe their blending, the more CPU time will be required
(see Sect. \ref{times}).
If the blending is not dramatic, it is safe to switch to the 
\textit{cells-on-objects} method
(i.e. put \texttt{fitting coo} in the parameter file). On the other
hand, if blending is severe this option would result in 
redundant fittings because cells would be enlarged to include
as many neighbours as possible, increasing the total computing time.
In this case, either stick to the whole image fitting, or 
(depending on the desired degree of accuracy) 
switch to the \textsc{tfit}-like cells fitting.
\item Spend some time in checking the output catalogue, e.g. considering
with caution fits relative to sources having
flags $>0$ and covariance indices larger than 1.
\end{itemize}

\end{appendix}

\end{document}